\documentclass[longauth,structabstract]{aa1} 
\usepackage{graphicx}
\usepackage{color}
\usepackage{txfonts}
\usepackage{longtable}
\usepackage{supertabular}
\usepackage{natbib}
\usepackage{soul}
\defcitealias{Umetsu+12}{U12}
\begin{document}

\definecolor{green}{rgb}{0,0.5,0}
\definecolor{grey}{rgb}{0.4,0.5,0.7}
\newcommand{\ab}[1]{\textcolor{red}{\bf Andrea: #1}}
\newcommand{\msun}{M_{\odot}}
\newcommand{\ks}{km~s$^{-1}$}
\newcommand{\rv}{r_{\Delta}}
\newcommand{\mv}{M_{\Delta}}
\newcommand{\rtwo}{r_{200}}
\newcommand{\rfive}{r_{500}}
\newcommand{\rtwofive}{r_{2500}}
\newcommand{\rume}{r_{\rm{200,U}}}
\newcommand{\mtwo}{M_{200}}
\newcommand{\ctwo}{c_{200}}
\newcommand{\rhor}{\rho(r)}
\newcommand{\br}{\beta(r)}
\newcommand{\cv}{c_{\Delta}}
\newcommand{\rs}{r_{-2}}
\newcommand{\rh}{r_{\rm H}}
\newcommand{\rb}{r_{\rm B}}
\newcommand{\ri}{r_{\rm I}}
\newcommand{\rc}{r_{\rm c}}
\newcommand{\rn}{r_{\nu}}
\newcommand{\rr}{r_{\rho}}
\newcommand{\ra}{r_{\beta}}
\newcommand{\slos}{\sigma_{\rm{los}}}
\newcommand{\qrr}{Q_{\rm r}(r)}
\newcommand{\nr}{\nu(r)}
\newcommand{\vrf}{{\rm v}_{{\rm rf}}}

\title{CLASH-VLT: The mass, velocity-anisotropy, and pseudo-phase-space
  density profiles of the $z=0.44$ galaxy cluster MACS~J1206.2-0847
  \thanks{Based in large part on data collected at the ESO VLT
    (prog.ID 186.A-0798), at the NASA HST, and at the NASJ Subaru
    telescope}}

\author{A. Biviano\inst{\ref{ABi}} 
\and P. Rosati\inst{\ref{PRo}}
\and I. Balestra\inst{\ref{AMe},\ref{ABi}}
\and A. Mercurio\inst{\ref{AMe}} 
\and M. Girardi\inst{\ref{MGi},\ref{ABi}} 
\and M. Nonino\inst{\ref{ABi}} 
\and C. Grillo\inst{\ref{CGr}}
\and M. Scodeggio\inst{\ref{MSc}}
\and D. Lemze\inst{\ref{DLe}} 
\and D. Kelson\inst{\ref{DKe}}
\and K. Umetsu\inst{\ref{KUm}}
\and M. Postman\inst{\ref{MPo}}
\and A. Zitrin\inst{\ref{MBa}}
\and O. Czoske\inst{\ref{OCz}}
\and S. Ettori\inst{\ref{SEt},\ref{MMe}}
\and A. Fritz\inst{\ref{MSc}}
\and M. Lombardi\inst{\ref{MLo}}
\and C. Maier\inst{\ref{OCz}}
\and E. Medezinski\inst{\ref{EMe}}
\and S. Mei\inst{\ref{SMe1},\ref{SMe2}}
\and V. Presotto\inst{\ref{MGi}}
\and V. Strazzullo\inst{\ref{VStr}}
\and P. Tozzi\inst{\ref{PTo}}
\and B. Ziegler\inst{\ref{OCz}}
\and M. Annunziatella\inst{\ref{MGi},\ref{ABi}}
\and M. Bartelmann\inst{\ref{MBa}}
\and N. Benitez\inst{\ref{NBe}}
\and L. Bradley\inst{\ref{MPo}}
\and M. Brescia\inst{\ref{AMe}}
\and T. Broadhurst\inst{\ref{TBr}}
\and D. Coe\inst{\ref{MPo}}
\and R. Demarco\inst{\ref{RDe}}
\and M. Donahue\inst{\ref{MDo}}
\and H. Ford\inst{\ref{DLe}}
\and R. Gobat\inst{\ref{RGo}} 
\and G. Graves\inst{\ref{GGr1},\ref{GGr2}}
\and A. Koekemoer\inst{\ref{MPo}}
\and U. Kuchner\inst{\ref{OCz}}
\and P. Melchior\inst{\ref{PMe}}
\and M. Meneghetti\inst{\ref{SEt},\ref{MMe}}
\and J. Merten\inst{\ref{LMo}}
\and L. Moustakas\inst{\ref{LMo}}
\and E. Munari\inst{\ref{MGi},\ref{ABi}}
\and E. Reg\H{o}s\inst{\ref{ERe}}
\and B. Sartoris\inst{\ref{MGi}}
\and S. Seitz\inst{\ref{SSe1},\ref{SSe2}}
\and W. Zheng\inst{\ref{DLe}}
}

\offprints{A. Biviano, biviano@oats.inaf.it}

\institute{INAF/Osservatorio Astronomico di Trieste, via G. B. Tiepolo 11, 
I-34131, Trieste, Italy\label{ABi} \and
ESO-European Southern Observatory, D-85748 Garching bei M\"unchen, Germany\label{PRo} \and
INAF/Osservatorio Astronomico di Capodimonte, Via Moiariello 16 I-80131 Napoli, Italy\label{AMe} \and
Dipartimento di Fisica, 
Univ. degli Studi di Trieste, via Tiepolo 11, I-34143 Trieste, Italy\label{MGi} \and
Dark Cosmology Centre, Niels Bohr Institute, University of Copenhagen,
Juliane Maries Vej 30, 2100 Copenhagen, Denmark\label{CGr} \and
INAF/IASF-Milano, via Bassini 15, 20133 Milano, Italy\label{MSc} \and
Department of Physics and Astronomy, The Johns Hopkins University, 3400 North Charles Street, Baltimore, MD 21218, USA\label{DLe} \and
Observatories of the Carnegie Institution of Washington, Pasadena, CA 91 101, USA\label{DKe} \and
Institute of Astronomy and Astrophysics, Academia Sinica, P. O. Box 23-141, Taipei 10617, Taiwan\label{KUm} \and
Space Telescope Science Institute, 3700 San Martin Drive, Baltimore, MD 21218, USA\label{MPo} \and
 Institut für Theoretische Astrophysik, Zentrum für Astronomie, Universit\"at Heidelberg, Philosophenweg 12, D-69120 Heidelberg, Germany\label{MBa} \and
University of Vienna, Department of Astrophysics, T\"urkenschanzstr. 17, 1180 Wien, Austria\label{OCz} \and
INAF/Osservatorio Astronomico di Bologna, via Ranzani 1, I-40127 Bologna, Italy\label{SEt} \and
INFN, Sezione di Bologna; Via Ranzani 1, I-40127 Bologna, Italy\label{MMe}  \and
Dipartimento di Fisica, Universitá degli Studi di Milano, via Celoria 16, I-20133 Milan, Italy\label{MLo} \and
Department of Physics and Astronomy, The Johns Hopkins University, 3400 North Charles Street, Baltimore, MD 21218, USA\label{EMe} \and
GEPI, Paris Observatory, 77 Avenue Denfert Rochereau, F-75014 Paris, France\label{SMe1} \and
University Denis Diderot, 4 Rue Thomas Mann, F-75205 Paris, France\label{SMe2} \and
CEA Saclay, Orme des Merisiers, F-91191 Gif sur Yvette, France\label{VStr} \and
INAF/Osservatorio Astrofisico di Arcetri, Largo E. Fermi 5, 50125 Firenze, Italy\label{PTo}
\and
Instituto de Astrof\'{\i}sica de Andaluc\'{\i}a (CSIC), C/Camino Bajo de Hu\'etor 24, Granada 18008, Spain\label{NBe} \and
Department of Theoretical Physics, University of the Basque Country, P. O. Box 644, 48080 Bilbao, Spain\label{TBr} \and
Department of Astronomy, Universidad de Concepcion, Casilla 160-C, Concepcion, Chile \label{RDe} \and
Department of Physics and Astronomy, Michigan State University, East Lansing, MI 48824, USA\label{MDo} \and
Laboratoire AIM-Paris-Saclay, CEA/DSM-CNRS, Université Paris Diderot, Irfu/Service d'Astrophysique, CEA Saclay, Orme des Merisiers, F-91191 Gif sur Yvette, France\label{RGo} \and
Department of Astronomy, University of California, Berkeley, CA, USA\label{GGr1} \and
Department of Astrophysical Sciences, Princeton University, Princeton, NJ, USA\label{GGr2} \and
Department of Physics, The Ohio State University, Columbus, OH, USA \label{PMe} \and
Jet Propulsion Laboratory, California Institute of Technology, 4800 Oak Grove Dr, Pasadena, CA 91109, USA\label{LMo} \and
European Laboratory for Particle Physics (CERN), CH-1211, Geneva 23, Switzerland\label{ERe} \and
University Observatory Munich, Scheinerstrasse 1, D-81679 M\"unchen, Germany\label{SSe1} \and
Max-Planck-Institut f\"ur extraterrestrische Physik, Postfach 1312, Giessenbachstr., D-85741 Garching, Germany \label{SSe2}
}
 
\date{}

\abstract{}{We constrain the mass, velocity-anisotropy, and
  pseudo-phase-space density profiles of the $z=0.44$ CLASH cluster
  \object{MACS~J1206.2-0847}, using the projected phase-space
  distribution of cluster galaxies in combination with gravitational
  lensing.}{We use an unprecedented data-set of $\simeq 600$ redshifts
  for cluster members, obtained as part of a VLT/VIMOS large program,
  to constrain the cluster mass profile over the radial range
  $\sim$0--5 Mpc (0--2.5 virial radii) using the MAMPOSSt and Caustic
  methods. We then add external constraints from our previous
  gravitational lensing analysis.  We invert the Jeans equation to
  obtain the velocity-anisotropy profiles of cluster members. With the
  mass-density and velocity-anisotropy profiles we then obtain the
  first determination of a cluster pseudo-phase-space density
  profile.} {The kinematics and lensing determinations of the cluster
  mass profile are in excellent agreement.  This is very well fitted
  by a NFW model with mass $\mtwo=(1.4 \pm 0.2) \times 10^{15} \,
  \msun$ and concentration $\ctwo=6 \pm 1$, only slightly higher than
  theoretical expectations. Other mass profile models also provide
  acceptable fits to our data, of (slightly) lower (Burkert,
  Hernquist, and Softened Isothermal Sphere) or comparable (Einasto)
  quality than NFW.  The velocity anisotropy profiles of the passive
  and star-forming cluster members are similar, close to isotropic
  near the center and increasingly radial outside.  Passive cluster
  members follow extremely well the theoretical expectations for the
  pseudo-phase-space density profile and the relation between the
  slope of the mass-density profile and the velocity
  anisotropy. Star-forming cluster members show marginal deviations
  from theoretical expectations.}{This is the most accurate
  determination of a cluster mass profile out to a radius of 5 Mpc,
  and the only determination of the velocity-anisotropy and
  pseudo-phase-space density profiles of both passive and star-forming
  galaxies for an individual cluster. These profiles provide
  constraints on the dynamical history of the cluster and its
  galaxies. Prospects for extending this analysis to a larger cluster
  sample are discussed.}

\keywords{Galaxies: clusters: individual: MACS~J1206.2-0847, Galaxies: kinematics and dynamics, Galaxies: evolution, Cosmology: dark matter}

\titlerunning{CLASH cluster mass and velocity anisotropy profiles}
\authorrunning{A. Biviano et al.}

\maketitle

\section{Introduction}
\label{s:intro}
Clusters of galaxies are excellent cosmological natural
laboratories. They are the most massive systems in dynamical
equilibrium, and are thus extremely sensitive and effective
cosmological probes, especially through the study of the cluster mass
function \citep[e.g.][and references herein]{KB12}. These systems are
believed to be dominated by dark matter \citep[DM
  hereafter,][]{Zwicky33}, so their internal mass distribution can in
principle be used to distinguish between DM and alternative theories
of gravity \citep[e.g.][]{Clowe+06b}, or to constrain the intrinsic
physical properties of DM
\citep[e.g.][]{ABG02,Markevitch+04,KBM04,SDR11}.

According to Cold DM cosmological numerical simulations, the radial
mass distribution of DM halos is universal, and their mass density
profiles can be characterized by a simple function
of the radial distance \citep[NFW model hereafter;][]{NFW96,NFW97}, at
least out to the virial radius\footnote{The radius $\rv$ is the radius
  of a sphere with mass overdensity $\Delta$ times the critical
  density at the cluster redshift. Throughout this paper we refer to
  the $\Delta=200$ radius as the 'virial radius', $\rtwo$.},
$\rtwo$. The NFW model parameters are the virial
radius $\rtwo$, and the scale radius $\rs$, that is the radius where
the logarithmic derivative of the mass density profile $\gamma \equiv
{\rm d}\ln \rho / {\rm d}\ln r =-2$. Equivalently, the NFW model can
be characterized by the related parameters, the virial
mass\footnote{The mass $\mv$ is directly connected to $\rv$ via $\mv
  \equiv \Delta \, H_z^2 \, \rv^3/(2 \, G)$, where $H_z$ is the Hubble
  constant at the redshift, $z$, of the halo. Throughout this paper we
  refer to the $\Delta=200$ mass as the 'virial mass', $\mtwo$.}
$\mtwo$, and the concentration $\ctwo \equiv \rtwo/\rs$.
An even
better fit to the density profile of cosmological halos can be
  obtained using the \citet{Einasto65} model \citep{Navarro+04}.
Observations have confirmed that the universal NFW model provides
adequate fit to the mass distribution of clusters
\citep[e.g.][]{Carlberg+97-mprof,GDK99,vanderMarel+00,KCS02,BG03,RGKD03,Kneib+03,KBM04,APP05,Broadhurst+05,Umetsu+11,Oguri+12,Okabe+13}.

Many studies have attempted to explain the NFW-like shape of the mass
density profile of cosmological halos, and why this shape is
universal, even if universality is still a debated issue
\citep[e.g.][]{Ricotti03,TKGK04,Merritt+06,RPV07}. While some studies have
found the shape of halo density profiles to depend on cosmology
\citep[e.g.][]{SCO00,Thomas+01,SMGH07}, others have not
\citep{HJS99_MN,WW09}. A general consensus is growing that the
universal NFW-like shape, at least in the central regions, is the
result of the initial, fast assembly phase of halos
\citep{HJS99,Manrique+03,ADK04,TKGK04,LMKW06,ElZant08,WW09,LC11},
characterized by dynamical processes such as violent and collective
relaxation, and phase and chaotic mixing \citep[][and references
  therein]{Henon64,LyndenBell67,Merritt05,Henriksen06}. The following
slower accretion phase may be responsible for the outer slope of the
density profile \citep{TKGK04,LMKW06,Hiotelis06}. Halos would obtain
the same, universal density profile independently of details about
their collapse \citep{ElZant08,WW09} and subsequent merger histories
\citep{Dehnen05,KZK06,ElZant08,WW09,SVMS12}.

It has been argued by \citet{TN01} that the NFW-like shape is strictly
related to the power-law radial behavior of the pseudo-phase-space
density profiles of halos identified in cosmological numerical
simulations, $Q(r) \equiv \rho/\sigma^3 \propto r^{-\alpha}$ with
$\alpha=-1.875$. This power-law behavior of $Q(r)$ is obeyed by a
variety of self-gravitating collisionless systems in equilibrium, not
necessarily formed as the result of hierarchical accretion processes,
and this suggests that it is a generic result of the collisionless
collapse, probably induced by violent relaxation
\citep{Austin+05,Barnes+06}.  A similar power-law behavior is also
obtained for $\qrr$, where the total velocity dispersion $\sigma$ is
replaced with its radial component, $\sigma_{\rm r}$ \citep{DML05}.

The power-law behavior may however not hold at all radii
\citep{Schmidt+08,Ludlow+10} and depending on the virialization state
of the system, departure from power-law may start already close to the
center, or, for more virialized halos, near the virial radius
\citep{Ludlow+10}. In any case, the relation is surprisingly similar
to the self-similar solution of \citet{Bertschinger85} for secondary
infall onto a spherical perturbation, even if the reason for this
similarity remains unexplained.

\citet{DML05} have shown that the shape of the density profiles of
cosmological halos follows analytically from the power-law behavior of
$Q(r)$ if the system obeys the Jeans equation of dynamical equilibrium
\citep{BT87}, and if a linear $\gamma$-$\beta$ relation holds, with
\begin{equation}
\br = 1 - {\sigma_\theta^2(r) + \sigma_\phi^2(r)  \over
  2\,\sigma_r^2(r)} = 1 - {\sigma_\theta^2(r) \over \sigma_r^2(r)}  
\label{e:beta}
\end{equation}
where $\sigma_\theta, \sigma_\phi$ are the two tangential components,
and $\sigma_r$ the radial component, of the velocity dispersion, and
the last equivalence is obtained in the case of spherical symmetry.
The existence of such a linear $\gamma$-$\beta$ relation has been found 
by \citet{HM06} to hold in a variety of halos
extracted from numerical simulations,
\begin{equation}
\br = -0.15 -0.19\, \gamma(r) \ .
\label{e:betaHM}
\end{equation}
The reality of this relation has been questioned by \citet{Navarro+10}
and \citet{Lemze+12} and yet some relation does seem to exist between
the shape of a halo mass density profile and the orbital properties of
the halo constituents \citep[see
  also][]{AE06,HMS06,ISNM06,HJS10,VHBD11}.  For a NFW-like density
profile, the $\gamma$-$\beta$ relation would imply isotropic orbits
($\beta \approx 0$) near the center, and more radially anisotropic
orbits ($\beta>0$) outside, as observed in DM halos. The radius where
$\br$ departs from isotropy, $\ra$, is then naturally related to the
characteristic scale length $\rs$ of the DM density profile
\citep{BWBD05,Bellovary+08}. A relation between $\rs$ and $\ra$ has
indeed been found in numerically simulated halos \citep{BWBD07,MBM10}.

Like the power-law behavior of $Q(r)$, also the $\gamma$-$\beta$
relation might be related to the halo formation process.
Isotropization of orbits may result from fluctuations in the
gravitational potential during the fast-accretion phase characterized
by major mergers, i.e.  a sort of violent or chaotic relaxation
\citep{LMKW06,LC11}.  The subsequent slow, gentle phase of mass
accretion is unable to isotropize orbits and as a consequence the
external, more recently accreted material would tend to move on more
radially elongated orbits. Another process capable of generating
isotropic orbits near the center of halos from an initial distribution
of radial orbits is the radial orbit instability (ROI hereafter). ROI
occurs when particles in precessing elongated loop orbits experience a
torque due to a slight asymmetry, that causes them to lose some
angular momentum and move towards the system center \citep[see,
  e.g.,][]{Bellovary+08}. ROI continues even after the halo has
virialized \citep{BWBD07}.

So far we have seen that the {\em shapes} of the mass density and
velocity anisotropy profiles seem to carry information on the
formation processes of cosmological halos but not on the cosmological
model. The latter might however be constrained by the relation between
the two parameters of the mass density profile, $\ctwo$ and $\mtwo$,
the so-called concentration-mass relation ($cMr$ hereafter).  In fact,
the halo concentration is determined by the mass fraction accreted
into the cluster during the initial fast phase \citep{LMKW06} so
$\ctwo$ and $\mtwo$ identify to a large extent the formation redshift
of a halo \citep[see, e.g.,][]{Gao+08,GTS12}. Observing the $cMr$ at
different redshifts can therefore be used to constrain cosmological
models \citep[see, e.g.,][]{HJS99_MN,Dolag+04,WT12}. For example, it
has been found that the $cMr$ has opposite slopes in Cold and Hot DM
cosmologies \citep{WW09}, while in dark-energy-dominated Warm
DM models the $cMr$ is not monotonous but characterized
by a turnover point at group mass scales  \citep{SSMM12}.

At present there is some tension between the observed $cMr$
\citep[e.g.][]{Lokas+06,RD06,Buote+07,SA07,Biviano08,Ettori+10,Okabe+10,Oguri+12,Newman+12}
and that obtained in $\Lambda$CDM cosmological simulations
\citep[e.g.][]{NFW97,Bullock+01,Duffy+08,Gao+08,KTP11,MMGD11,GMEM12,BHHV13},
particularly at the low mass end (galaxy groups). The use of the $cMr$
for discriminating among different cosmological models is however
somewhat hampered by our ignorance of baryon-related physical
processes that can change halo concentrations, also as a function of
halo mass \citep[e.g.][]{ElZant+04,GKKN04,BL10,DelPopolo10,Fedeli12}.
\citet{RBEMM13} have shown that the effect of baryons is not
enough to reconcile the observed and simulated $cMr$. Efficient radiative
cooling and weak feedback are needed to reconcile the observed and
simulated $cMr$ on the scale of galaxy groups, but this comes at the 
price of creating tension with other observables, such as the stellar
mass fraction \citep{Duffy+10}.

The above theoretical considerations about the universality, the
shape, and the origin of cluster mass profiles need to be tested
observationally.  Determining cluster mass profiles is however not a
simple task. Traditionally, this has been done using cluster galaxies
as tracers of the gravitational potential \citep[e.g.][and references
  therein]{KG82,TW86,vanderMarel+00,BG03,Biviano00} -- this technique
has allowed the first discovery of dark matter \citep{Zwicky33}. The
intra-cluster gas has been used as tracer of the gravitational
potential since the advent of X-ray astronomy
\citep[e.g.][]{MIC77,FJ82,Fabricant+86,BHB92,EDGM02}. Cluster masses
and mass profiles can also be measured using the Sunyaev-Zeldovich
\citep{SZ70b} effect
\citep[e.g.][]{Pointecouteau+99,Grego+00,LaRoque+03,Muchovej+07}, but
perhaps the most direct way is by exploiting the gravitational
distortion effects of the cluster potential on the apparent shapes of
background galaxies
\citep[e.g.][]{WSGW89,MFK93,Squires+96,STE02,DHS03,Zitrin+11b} as
first suggested by \citet{Zwicky37}.

Using different methods to determine cluster mass profiles is
fundamental since different methods suffer from different systematics.
For instance, X-ray determinations of cluster masses tend to be
underestimated if bulk gas motions and the complex thermal structure
of the Intra-Cluster Medium (ICM) are ignored
\citep{RTM04,Rasia+06,LKN09,Molnar+10,CLF11}.  Cluster triaxiality and
orientation effects tend to bias the mass profile estimates obtained
by gravitational lensing \citep[e.g.][]{Meneghetti+11,BK11,FH12} and
by cluster galaxy kinematics \citep{Cen97,Biviano+06}.  Comparing
different mass profile determinations can therefore help assessing the
contribution of non-thermal pressure to the ICM and the elongation
along the line-of-sight \citep[e.g.][]{ML12,SEB12}. If systematics are
well under control, the comparison of independent determinations of
cluster mass profiles from gravitational lensing and the kinematics of
cluster members can shed light on the very nature of DM \citep{FV06,SDR11}.

While different methods can be used to constrain a cluster mass
profile, direct determination of its velocity-anisotropy profile
$\br$ can only be achieved by using cluster galaxies as tracers
of the gravitational potential
\citep{KG82,KS83,MP86,SEG88,NK96,Biviano+97,Carlberg+97-equil,ABM98,Mahdavi+99,LM03,BK04,MG04,Benatov+06,Lokas+06,HL08,Adami+09,BP09,Lemze+09,WL10}. 

In this paper we present a new determination of the mass and velocity
anisotropy profiles of a massive, X-ray selected cluster at redshift
$z=0.44$, largely based on spectroscopic data collected at ESO
VLT. These data have been collected within the ESO Large Programme
186.A-0798 ``Dark Matter Mass Distributions of Hubble Treasury
Clusters and the Foundations of $\Lambda$CDM Structure Formation
Models'' (P.I. Piero Rosati). This is an ongoing spectroscopic
follow-up of a subset of 14 clusters from the ``Cluster Lensing And
Supernova survey with Hubble'' \citep[CLASH,][]{Postman+12}. The
CLASH-VLT Large Programme is aimed at obtaining redshift measurements
for 400--600 cluster members and 10--20 lensed multiple images in each
cluster field. We combine our cluster mass profile determination based
on spectroscopic data for member galaxies, with independent mass
profile determinations obtained from the strong and weak gravitational
lensing analyses of, respectively, \citet{Zitrin+12} and \citet[][U12
  hereafter]{Umetsu+12}. The combined power of the excellent imaging
and spectroscopic data allows us to determine the mass profile for a
single cluster to an unprecedented accuracy and free of systematics
over the radial range $\sim$0--5 Mpc (corresponding to 0--2.5 virial
radii).  The cluster mass profile so obtained is then used to
determine the velocity anisotropy profiles, $\br$ of both the passive
and star-forming cluster galaxy populations, for the first time for an
individual cluster, thanks to the large sample of spectroscopic
redshifts.  This is the highest-redshift determination of $\br$ for an
individual cluster so far. The mass profile and $\br$ determinations
are then used to determine (for the first time ever for a real galaxy
cluster) the pseudo-phase-space density profiles $Q(r)$ and $\qrr$,
and the $\gamma$-$\beta$ relation.

The paper is organized as follows. In Section~\ref{s:data} we describe
the data sample, and the identification of cluster members. We
determine the cluster mass profile in Section~\ref{s:Mprof} and
compare our results to theoretical expectations for the $cMr$. We determine
the cluster velocity anisotropy profile in Section~\ref{s:beta}.  In
Section~\ref{s:qr} we test observationally the theoretical $Q(r)$,
$\qrr$, and $\gamma$-$\beta$ relation. We discuss our results in
Section~\ref{s:disc} and provide our conclusions in
Section~\ref{s:conc}. In Appendix~\ref{s:syst} we show that our
results are robust vs. different choices of the method for cluster
members identification. In Appendix~\ref{s:compMr} we compare our
results for the cluster mass with previous, less accurate results from
the literature.

Throughout this paper, we adopt 
$H_0=70$ km~s$^{-1}$~Mpc$^{-1}$, $\Omega_0=0.3$, $\Omega_\Lambda=0.7$.
At the cluster redshift, 1 arcmin corresponds to 0.34 Mpc. Magnitudes
are in the AB system.

\section{The data sample}
\label{s:data}
The cluster MACS~J1206.2-0847 was observed in 2012 as part of the ESO
Large Programme 186.A-0798 using VIMOS \citep{LeFevre+03} at the ESO
VLT. The VIMOS data were acquired using four separate pointings, each with
a different quadrant centered on the cluster core.  A total of 12 masks were
observed (8 LR-Blue masks and 4 MR masks), and each mask was observed for
either 3 or $4 \times 15$ minutes, for a total of 10.7 hours exposure
time. The LR-Blue masks cover the spectral range 370-670 nm with a
resolution R=180, while the MR masks cover the range 480-1000 nm with
a resolution R=580.

We used VIPGI \citep{Scodeggio+05} for the spectroscopic data reduction.
We assigned a Quality Flag (QF) to each redshift, which qualitatively
indicates the reliability of a redshift measurement. We define four
redshift quality classes: ``secure'' (QF=3), ``likely'' (QF=2),
``insecure'' (QF=1), and ``based on a single-emission-line''
(QF=9). To assess the reliability of these four quality classes we
compared pairs of duplicate observations having at least one secure
measurement.  Thus, we could quantify the reliability of each quality
class as follows: redshifts with QF=3 are correct with a probability
of $>99.99$\%, QF=9 with $\sim 92$\% probability, QF=2 with $\sim
75$\% probability, and QF=1 with $<40$\% probability. We do not
consider QF=1 redshifts in this paper.

Additional spectra were taken from \citet{Lamareille+06} (3 objects),
\citet{Jones+04} (1), \citet{Ebeling+09} (25), and Daniel Kelson (21
observed with IMACS-GISMO at the Magellan telescope, private
communication). Archival data from the programs 169.A-0595 (PI: Hans
B\"ohringer; 5 LR-Blue masks) and 082.A-0922 (PI: Mike Lerchster, 1
LR-Red mask), for 952 spectra in the cluster field were reduced
following the same procedure used for our new CLASH-VLT data, using the
appropriate calibrations.

The final data-set contains 2749 objects with reliable redshift
estimates, of which 2513 have $z>0$, 18\% of them obtained in MR
mode. Repeated measurements of the same spectra were used to estimate
the average error on the radial velocities, 75 (153) \ks for the
spectra observed with the MR (LR, respectively) grism. The average
error is sufficiently small not to affect our dynamical analysis,
given the large velocity dispersion of the cluster.  Full details on
the spectroscopic sample observations and data-reduction will be given
in Rosati et al. (in prep.).

Photometric data were derived from Suprime-Cam observations at the
prime focus of the Subaru telescope, in five bands 
\citepalias[$BVR_cI_cz'$, see][]{Umetsu+12}.  Full details on the derivation of the
photometric catalog used in this paper will be given in Mercurio et
al. (in prep.).

\subsection{Cluster membership: the spectroscopic sample}
\label{ss:members}
Several methods exist to identify cluster members in a spectroscopic
data-set \citep[see][and references therein]{Wojtak+07}. Most of them
are based on the location of galaxies in projected
phase-space\footnote{We call $R$ (resp. $r$) the projected
  (resp. 3D) radial distance from the cluster center (we
  assume spherical symmetry in the dynamical analyses).  The
  rest-frame velocity is defined as $\vrf \equiv c \,
  (z-\overline{z})/(1+\overline{z})$, where $\overline{z}$ is the mean
  cluster redshift, redefined at each new iteration of the membership
  determination.}, $R, \vrf$. For the cluster center we choose the
position of the Brightest Cluster Galaxy (BCG,
$\alpha_{\mathrm{J2000}} = 12^{\mathrm{h}}06^{\mathrm{m}}12\fs15,
\delta_{\mathrm{J2000}} = -8\degr 48\arcmin 3\farcs4$ ). The BCG
position practically coincides with the X-ray peak position and the
center of mass determined by the gravitational lensing analysis
\citepalias{Umetsu+12}, as all these three positions are within 13 kpc
from each other.

\begin{figure}
\begin{center}
\begin{minipage}{0.5\textwidth}
\resizebox{\hsize}{!}{\includegraphics{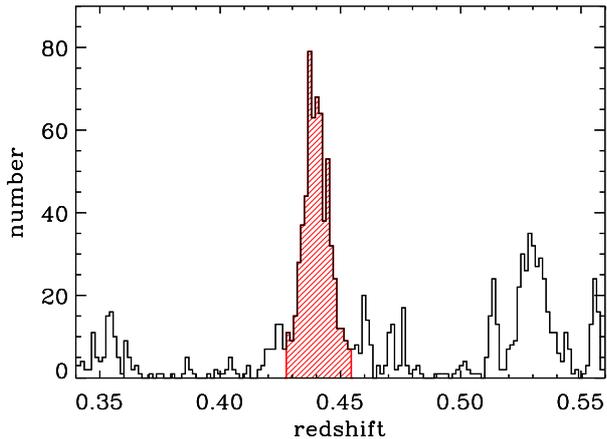}}
\end{minipage}
\end{center}
\caption{Histogram of redshifts in the cluster area. The red, hatched
  histogram shows the main cluster peak identified by the P+G method.}
\label{f:peak}
\end{figure}

Here we consider two methods to assign the cluster membership, the
method of \citet{Fadda+96}, that we call 'P+G' (Peak+Gap), and the
'Clean' method of \citet{MBB13}. The two methods are very different;
in particular, unlike the Clean method, the P+G method does not make
any assumption about the cluster mass profile. In both methods the
main peak in the $z$-distribution is identified. For this, P+G uses an
algorithm based on adaptive kernels \citep{Pisani93}, and Clean uses
the weighted gaps in the velocity distribution.  After the main peak
identification (shown in Fig.~\ref{f:peak}) P+G considers galaxies in
moving, overlapping radial bins to reject those that are separated
from the main cluster body by a sufficiently large velocity gap (we
choose $\Delta \vrf=800$ \ks).  The Clean method uses a robust
estimate of the cluster line-of-sight velocity dispersion, $\slos$, to
guess the cluster mass using a scaling relation. It then adopts the
NFW profile, the theoretical $cMr$ of \citet{MDvdB08}, and the velocity
anisotropy profile model of \citet{MBM10}, to predict $\slos(R)$ and
to iteratively reject galaxies with $\mid \vrf \mid > 2.7 \, \slos$ at
any radius.

\begin{figure}
\begin{center}
\begin{minipage}{0.5\textwidth}
\resizebox{\hsize}{!}{\includegraphics{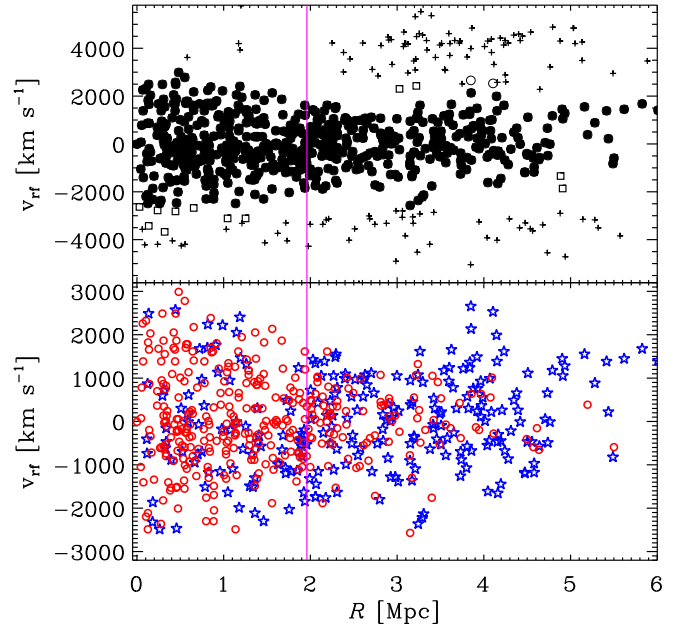}}
\end{minipage}
\end{center}
\caption{{\em Top panel:} Galaxies in the projected phase-space
  diagram, $R, \vrf$.  Black dots represent galaxies identified as
  cluster members by both the P+G and Clean algorithms. Open circles
  represent galaxies identified as cluster members by the P+G
  algorithm only.  Squares represent galaxies identified as cluster
  members by the Clean algorithm only.  Crosses represent non-cluster
  members.  {\em Bottom panel:} Cluster members selected with the P+G
  method in the projected phase-space diagram, $R, \vrf$. Red circles
  represent passive galaxies, blue stars represent SF galaxies.  In
  both panels the vertical (magenta) line indicates $\rume$, i.e. the
  $\rtwo$ value obtained by scaling the $\rv$ estimate of
  \citetalias{Umetsu+12} at $\Delta=200$, using their best-fit NFW
  profile.}
\label{f:rvm}
\end{figure}

In Fig.~\ref{f:rvm} (top panel) we show the $R, \vrf$ cluster diagram,
with the cluster members selected by the two methods. The P+G and
Clean method select 592 and 602 cluster membersd, respectively.  This
is one of the largest spectroscopic sample for members of a
single cluster, and the largest at $z>0.4$.  There are 590 members in
common between the two methods, meaning that only two P+G members are
not selected by the Clean method, while 12 Clean members are not
selected by the P+G method. Given that the two methods are very
different, these differences can be considered quite marginal. Since
one of our aims is to determine the cluster mass profile, we prefer to
base our analysis on the sample of members defined with the P+G
method, because, at variance with the Clean method, it requires no
{\em a priori} assumptions about the cluster mass profile. In
Appendix~\ref{s:syst} we show that our results are little affected if
we choose the Clean membership definition instead.

\begin{table}
\centering
\caption{Values of the line-of-sight velocity dispersions, $\slos$, and of the best-fit parameters of the galaxy number density profiles, $n(R)$.}
\label{t:slos}
\begin{tabular}{lccccc}
\hline 
Sample     &  $\slos$           & \multicolumn{3}{c}{$n(R)$} \\
           &                  & \multicolumn{2}{c}{scale radius $\rn$} & model \\
           & km~s$^{-1}$    &  \multicolumn{2}{c}{[Mpc]} & \\
           &                    & spec & spec+phot & \\ 
\hline
& & & \\
All        & $1087_{-55}^{+53}$ & $0.74_{-0.17}^{+0.10}$ & $0.63_{-0.09}^{+0.11}$ & pNFW \\
& & & & \\
Passive    & $1042_{-53}^{+50}$ & $0.61_{-0.11}^{+0.15}$ & $0.56_{-0.08}^{+0.12}$ & pNFW  \\
& & & & \\
SF         & $1144_{-58}^{+55}$ & $0.61_{-0.17}^{+0.20}$ & $0.57_{-0.17}^{+0.24}$ & King \\
& & & & \\
\hline
\end{tabular}
\tablefoot{The scale radius best-fit values are given for two
  selections of members; 'spec' refers to the purely spectroscopic
  selection (also used for the determination of $\slos$), 'spec+phot'
  to the combined spectroscopic and photometric selection (for details
  see Sect.~\ref{ss:compl}). The models used for $n(R)$ are the
  projected NFW ('pNFW'), and \citet{King62_clusters}'s ('King').}
\end{table}

Using the P+G members, we estimate the cluster
mean\footnote{Throughout this paper we use the robust biweight
  estimator for computing averages and dispersions \citep{BFG90}, and
  eqs.(15) and (16) in \citet{BFG90} for computing their
  uncertainties.} redshift $\overline{z}=0.43984 \pm 0.00015$. The
cluster velocity dispersion is given in Table~\ref{t:slos} with
1~$\sigma$ errors. 

\begin{figure}
\begin{center}
\begin{minipage}{0.5\textwidth}
\resizebox{\hsize}{!}{\includegraphics{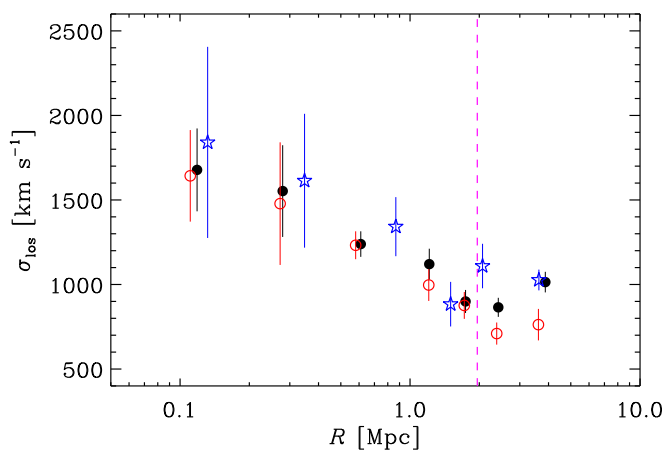}}
\end{minipage}
\end{center}
\caption{Line-of-sight velocity dispersion profiles of cluster members
  (using the P+G identification method). Black filled dots: all
  galaxies; red circles: passive galaxies; blue stars: SF
  galaxies. 1~$\sigma$ error bars are shown. The vertical magenta
  dashed line indicates $\rume$.}
\label{f:slos}
\end{figure}

Since the velocity distribution of late-type/blue/active galaxies
  in clusters is different from that of early-type/red/passive
  galaxies and characterized by a larger $\slos$ \citep[at least in
  nearby
  clusters;][]{Tammann72,MD77,Sodre+89,Biviano+92,Biviano+97,Carlberg+97-equil,Einasto+10},
it is worth considering a subsample of red/passive galaxies for an
estimate of $\slos$ and, thereby, $\rtwo$. To select a subsample of
passive galaxies we use their location in a color-color plot,
requiring $(m_V-m_I) \leq -10.47 + 5.5 \, (m_B-m_R)$.  This
color-color selection separates two subsamples of high-quality
spectrum galaxies showing spectroscopic features typical of a
passively-evolving stellar population and, separately, of ongoing
star-formation (for details see Mercurio et al., in prep.).

The velocity dispersions of passive and star-forming (SF hereafter)
galaxies are not significantly different (see Table~\ref{t:slos}).
This is also evident from the distribution of the two samples in the
$R, \vrf$ diagram (Fig.~\ref{f:rvm}, bottom panel) and from
the $\slos$ profiles shown in Fig.~\ref{f:slos}. In nearby clusters
there is more difference between the $\slos$ profiles of the
passive and SF galaxy populations, but this difference is known to
become less significant in higher-$z$ clusters \citep{BP09,BP10}.

We obtain a first estimate of the cluster $\mtwo$ and $\rtwo$ from the
$\slos$ estimate of the passive cluster members, following the method
of \citet{MBB13}. We assume that (i) the mass is distributed
according to the NFW model, (ii) the NFW concentration parameter is
obtained iteratively from the mass estimate itself using the
$cMr$ of \citet{MDvdB08}, and (iii) the
velocity anisotropy profile is that of \citet{ML05b} with a scale
radius identical to that of the NFW profile \citep[as found in
  cluster-mass halos extracted from cosmological numerical
  simulations, see][]{MBM10,MBB13}. The procedure is iterative and
uses the value of $\slos$ re-calculated at each iteration on the
members within $\rtwo$. We find $\mtwo=1.42 \times 10^{15} \msun$,
which corresponds to $\rtwo=1.98$ Mpc.  Since $\rtwo \propto \slos$,
the $\slos$ uncertainty implies a $\simeq 5$\% formal fractional
uncertainty on the $\rtwo$ estimate, and three times larger on
$\mtwo$. 

This determination of $\rtwo$ is based on the assumption that the
velocity distribution of passive cluster members is unbiased relative
to that of DM particles. Numerical simulations suggest that a bias
exists, albeit small \citep[e.g.][]{Berlind+03,Biviano+06,Munari+13},
so we must take this result with caution. The MAMPOSSt and Caustic
methods we will use in the following (see Sects.~\ref{ss:MAM} and
\ref{ss:CAU}) are unaffected by this possible systematics.

Our $\slos$-based $\rtwo$ value is very close to that obtained by
\citetalias{Umetsu+12} from a gravitational lensing analysis, 1.96
Mpc. We estimate this value using their best-fit NFW $\mv$ and $\cv$
values converted from their adopted $\Delta=131$ to $\Delta=200$ (we
do the same for $\cv$, see Table~\ref{t:NFW}).  Hereafter we refer to
\citetalias{Umetsu+12}'s value of $\rtwo$ as $\rume$.

\subsection{Completeness and number density profiles}
\label{ss:compl}
\begin{figure}
\begin{center}
\begin{minipage}{0.5\textwidth}
\resizebox{\hsize}{!}{\includegraphics{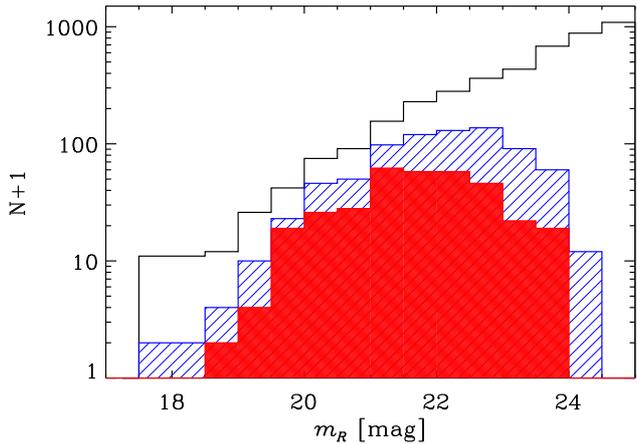}}
\end{minipage}
\end{center}
\caption{$R_c$-band number counts in the cluster virial region (within a
  radius $R \leq 1.96$ Mpc) for all photometric objects (black
  histogram), for objects with measured redshifts (hatched blue
  histogram), and for cluster spectroscopic members (filled red
  histogram).}
\label{f:comphisto}
\end{figure}

Our spectroscopic sample is not complete down to a given flux. This
can be seen in Fig.~\ref{f:comphisto} where we show the $R_c$-band
number counts in the cluster virial region ($R \leq 1.96$ Mpc), for
all photometric objects, for objects with measured redshifts, and for
cluster spectroscopic members (see Sect.~\ref{ss:members}).  Note that
the target selection in the spectroscopic masks is such to span a wide
color range, so that the resulting sample does not have any
appreciable bias against galaxies of a given type, which span from
early-type to actively star-forming. 

The incompleteness of the spectroscopic sample is not relevant for
that part of the dynamical analysis which is based on the velocity
distribution of cluster members. This distribution can be determined
at different radii even with incomplete samples, the only
effect of incompleteness being a modulation of the accuracy with which
the velocity distribution can be estimated at different radii.

The incompleteness of the spectroscopic sample can instead affect the
determination of the cluster projected number density profile, $n(R)$,
which converts to the 3D number density profile $\nr$ via the Abel
integral equation \citep{BT87}. The absolute normalization of the
galaxy number density profile $\nr$ is of no concern, however, for our
dynamical analysis, since it is only the logarithmic derivative of
$\nr$ that enters the Jeans equation \citep[see, e.g., eq. 4
  in][]{KBM04}. Only if the incompleteness of the sample is not the
same at all radii must we be concerned.

\begin{figure}
\begin{center}
\begin{minipage}{0.5\textwidth}
\resizebox{\hsize}{!}{\includegraphics{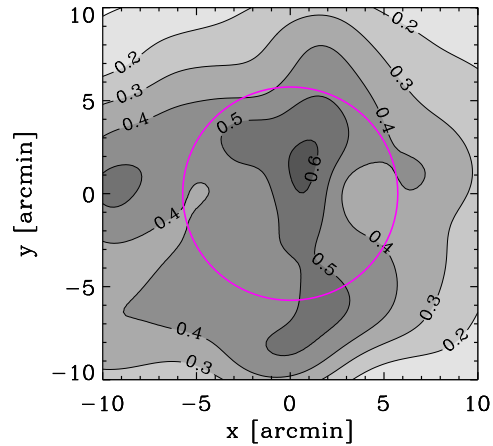}}
\end{minipage}
\end{center}
\caption{Spectroscopic completeness map. This is the ratio of two
  adaptive-kernel number density maps, one for all the objects with
  $z$, and the other for all the photometric objects, both within the
  magnitude range $18 \leq m_R \leq 23$. Contours are labeled with the
  completeness levels, and show that the spectroscopic completeness
  becomes slightly higher closer to the center. The magenta circle
  represents the virial region with radius $R \leq \rume$.}
\label{f:compmap}
\end{figure}

Our spectroscopic sample does have a mild radially-dependent
incompleteness. This is illustrated in Fig.~\ref{f:compmap} where we
show a spectroscopic-completeness map obtained as the ratio of two
adaptive-kernel maps of galaxy number densities, one for all the
objects with $z$, and the other for all the photometric objects. In
both cases we only consider objects within the magnitude range covered
by most of the spectroscopic cluster members, $18 \leq m_R \leq 23$.

\begin{figure}
\begin{center}
\begin{minipage}{0.5\textwidth}
\resizebox{\hsize}{!}{\includegraphics{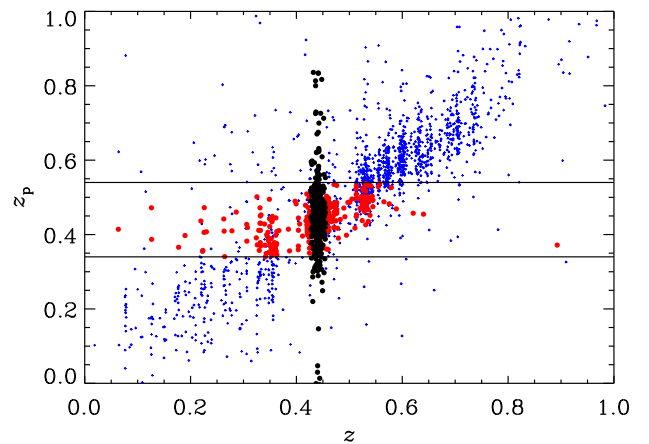}}
\end{minipage}
\end{center}
\caption{Photometric $z_{\rm p}$ vs. spectroscopic $z$ for the sample
  of galaxies with $z$ and $18 \leq m_R \leq 23$ in the cluster
  field. Spectroscopic cluster members are indicated with black dots,
  galaxies selected within the $0.34<z_{\rm p}<0.54$ range {\em and}
  within the chosen $m_R-m_I$ vs.  $m_B-m_V$ color-color cut (see
  text) are indicated with red (grey) dots. Galaxies outside the
  photometric and spectroscopic membership selection are indicated
  with blue crosses.}
\label{f:zpz}
\end{figure}

\begin{figure}
\begin{center}
\begin{minipage}{0.5\textwidth}
\resizebox{\hsize}{!}{\includegraphics{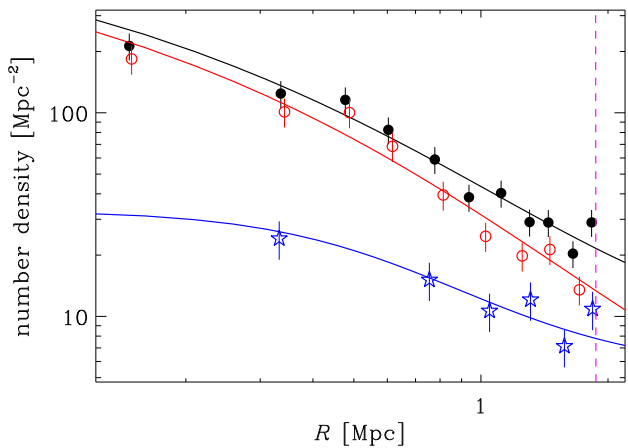}}
\end{minipage}
\end{center}
\caption{Projected galaxy number density profiles $n(R)$ (symbols with
  1~$\sigma$ error bars) and best-fits (solid lines) for the whole
  cluster population (black filled dots), for the population of
  passive cluster galaxies (red open dots), and for the population of
  SF cluster galaxies (blue stars). The best-fit models are pNFW for
  all and passive members, and the model of
  \citet{King62_clusters} for the SF members. A constant galaxy
  density background is added to all models.  The vertical magenta
  dashed line represents $\rume$.}
\label{f:nr}
\end{figure}

We need to know the radially-dependent completeness correction with an
adequate spatial resolution to correctly sample $\nr$ at small radii,
but increasing the spatial resolution comes at the price of increasing
the Poisson noise of the number counts on which we base our
completeness estimates. Given that within $\rume$ the
  spectroscopic completeness varies by less than $\sim 20$\%
(Fig.~\ref{f:compmap}) we can, to first approximation, ignore this
  mild radially-dependent incompleteness. We therefore determine the
  galaxy $n(R)$ directly from our spectroscopic sample of members
  within the virial radius and with magnitudes $18 \leq m_R \leq 23$.

 We fit the number density profile of the full sample of cluster
   members, and, separately, the profiles of the subsamples of passive
   and SF galaxies (defined in Section~\ref{ss:members}), using a
   Maximum Likelihood technique, which does not require radial binning
   of the data \citep{Sarazin80}.  We fit the data with either a
 projected NFW model \citep[pNFW hereafter;][]{Bartelmann96} or with a
 King model, $n(R) \propto 1/[1+(R/\rc)^2]$
 \citep{King62_clusters,Adami+98-7}.  The only free parameter in
   these fits is the scale radius.  The results are given in
 Table~\ref{t:slos}.  The pNFW model provides a better fit than
   the King model for the samples of all and passive members, while
   the King model is preferable to the pNFW model for the sample of SF
   galaxies. All fits are acceptable within the $46$\% confidence
   level, with reduced $\chi^2$ of 0.9, 0.6, and 0.3, for the
   populations of all, passive, and SF galaxies, respectively.

To assess the effect of unaccounted incompleteness bias in our
  estimates, we now check these results using a nearly complete sample of
galaxies. This is the sample of galaxies with available photometric
redshifts, $z_{\rm p}$. Note that we only use this photometric
  sample for the determination of $n(R)$.  Our dynamical analysis is
  entirely based on the spectroscopic sample.

The $z_{\rm p}$ have been obtained by a method based on neural
networks. In particular we used the MultiLayer Perceptron
\citep[MLP,][]{Rosenblatt57} with Quasi Newton learning rule. The MLP
architecture is one of the most typical feed-forward neural network
model. The term feed-forward is used to identify the basic behavior of
such neural models, in which the impulse is propagated always in the
same direction, e.g. from neuron input layers towards output layers,
through one or more hidden layers (the network brain), by combining
sums of weights associated to all neurons (except the input
layer). Quasi-Newton Algorithms (QNA) are an optimization of learning
rule, in particular they are variable metric methods for finding local
maxima and minima of functions \citep{Davidon91}. The model based on
this learning rule and on the MLP network topology is then called
MLPQNA \citep[for details on the method see][]{Brescia+13,Cavuoti+12}.

This method was applied to the whole data-set of $\sim$ 34,000 objects
with available and reliable $BVR_cI_cz'$-band magnitudes down to
$m_R=25.0$, following a procedure of network training and validation
based on the subsample of objects with spectroscopic redshifts. We
splitted the spectroscopic sample into two subsets, using as the
training set 80\% of the objects and as the validation set the
remaining 20\%. In order to ensure a proper coverage of the
parameter space we checked that the randomly extracted populations had
a spectroscopic distribution compatible with that of the whole
spectroscopic sample. Using subsamples of objects with
spectroscopically measured redshifts as training and validation sets
makes the estimated $z_{{\rm p}}$ insensitive to photometric
systematic errors (due to zero points or aperture corrections). In
this sense this method is more effective than classical methods based
on Spectral Energy Distribution fitting (see Mercurio et al., in
prep., for further details on our $z_{{\rm p}}$ estimates).

We must identify cluster members among the galaxies with $z_{{\rm p}}$
and without spectroscopic redshifts to ensure that the number
density profile we determine is a fair representation of what we would
have obtained using a complete sample of spectroscopic
members. We define the cluster membership by
requiring $0.34<z_{\rm p}<0.54$ to ensure low contamination by
foreground and background galaxies, and yet include most cluster
members (see Fig.~\ref{f:zpz}). In the effort to limit field
contamination we also apply the following color-color cuts, chosen by
inspecting the location of the spectroscopic members in the
color-color diagram:
\begin{eqnarray}
-0.09+0.52 \, (m_B-m_V)<m_R-m_I<0.21+0.52 \, (m_B-m_V) \nonumber \\
\mbox{for} \; 0.20<m_B-m_V<0.45 \ , \nonumber
\end{eqnarray}
\begin{eqnarray}
-0.09+0.52 \, (m_B-m_V)<m_R-m_I<0.36+0.52 \, (m_B-m_V)  \nonumber \\
\mbox{for} \; 0.45 \leq m_B-m_V<0.80 \ , \nonumber
\end{eqnarray}
\begin{eqnarray}
0.01+0.52 \, (m_B-m_V)<m_R-m_I<0.36+0.52 \, (m_B-m_V)  \nonumber \\
\mbox{for} \; 0.80 \leq m_B-m_V < 1.30. \nonumber
\end{eqnarray}
To maximize the number of objects with spectroscopic redshifts we
consider the magnitude range $18 \leq m_R \leq 23$. We then add to
this sample the spectroscopic members defined in
Sect.~\ref{ss:members}.  The combined sample of spectroscopic and
photometric members contains 1597 galaxies, of which 54\% are
photometrically selected.

The purity of the sample of photometrically-selected members can be
estimated based on the sample of galaxies with both spectroscopic and
photometric redshifts.  We define the purity ${\rm P} \equiv N_{pm
  \cap zm}/N_{pm \cap z}$, where $N_{pm \cap z}$ (respectively, $N_{pm
  \cap zm}$) is the number of galaxies with $z$ (respectively, the
number of spectroscopically confirmed cluster members) which are
selected as photometric members. We find ${\rm P}=0.64$.  The
color-color selection is useful to reduce the contamination,
especially by background objects. Had we not used the color-color
selection, the purity would have been lowered to 0.50. If we assume
the spectroscopic sample of members to have ${\rm P}=1$, the combined
sample of photometric and spectroscopic members has ${\rm P}=0.82$.

We fit the number density profiles of this complete sample of
  (photometrically- and spectroscopically-selected) cluster members, both for the full
  sample, and for the subsamples of passive and SF galaxies (defined
  in Section~\ref{ss:members}), within the virial radius, using
  the same Maximum Likelihood technique already used for the
  spectroscopic sample. As before we consider either a pNFW or a King
  model, but this time we add an additional constant background
density parameter in both models. The background density parameter is
needed because we expect that the photometric membership selection is
contaminated by non-cluster members. From the estimate of the purity
of the sample, we expect 18\% of the selected members to be
spurious, and this corresponds to 8 background galaxies
Mpc$^{-2}$ in our sample of photometrically-selected members, 3/4 of
which are SF galaxies. This value is very close to the density of
photometrically-selected members in the external cluster regions,
$4<R<5$ Mpc, where the field contamination of this sample is likely to
be dominant.

Once the background galaxy density parameter is fixed, the only
remaining free parameter in the fit is the scale radius. The results
of our fits are given in Table~\ref{t:slos} and displayed in
Fig.~\ref{f:nr}. The pNFW model provides a better fit than the King
model for the samples of all and passive members, while the King model
is preferable to the pNFW model for the sample of SF galaxies. All
fits are acceptable within the $69$\% confidence level, with reduced
$\chi^2$ of 1.1, 1.2, and 0.8, for the populations of all, passive,
and SF galaxies, respectively. These results are very similar to those
obtained using the spectroscopically-selected cluster members.

In Sect.~\ref{ss:MAM} we will use the $n(R)$ best-fits of the
whole cluster population within the MAMPOSSt method. We will
  consider both results listed in Table~\ref{t:slos} to check how
  much our dynamical results depend on the best-fit solution for the
  $n(R)$ scale radius.  

\section{The mass profile}
\label{s:Mprof}
\subsection{The MAMPOSSt method}
\label{ss:MAM}
The MAMPOSSt method \citep{MBB13} aims to determine the mass and
velocity anisotropy profiles of a cluster in parametrized form, by
performing a maximum likelihood fit of the distribution of galaxies in
projected phase space. MAMPOSSt does not postulate a shape for the
distribution function in terms of energy and angular momentum, and
does not suppose Gaussian line-of-sight velocity distributions, but
assumes a shape for the 3D velocity distribution (taken to be Gaussian
in our analysis). This method has been extensively tested using
cluster-mass halos extracted from cosmological simulations. It
assumes dynamical equilibrium, hence it should not be applied to data
much beyond the virial radius. Following the indications of
\citet{MBB13} we only consider data within $R \leq \rtwo$.
We also exclude the very inner region, within
0.05 Mpc, since it is dominated by the internal dynamics of the BCG,
rather than by the overall cluster \citep[see, e.g.,][]{BS06}. Our
MAMPOSSt analysis is therefore based on the sample of 330 cluster
members with $0.05 \leq R \leq \rume$. Of these, 250 are passive
galaxies (see Section~\ref{ss:members}).

The MAMPOSSt method requires parametrized models for the number
density, mass, and velocity anisotropy profiles -- $\nr$, $M(r)$,
$\br$, but there is no limitation in the possible choice of these
models. Since our spectroscopic data-set might suffer from (mild)
radial-dependent incompleteness, we prefer not to let
MAMPOSSt fit $\nr$ directly; rather, we use the de-projected $n(R)$
best-fit models obtained 
in Sect.~\ref{ss:compl} (see
Table~\ref{t:slos}). We refer to the scale radius of the number
density profile as $\rn$ in the following.

As for $M(r)$, we consider the following models:
\begin{enumerate}
\item the NFW model,
\begin{equation}
M(r)=\mtwo {\ln(1+r/\rs)-r/\rs \, (1+r/\rs)^{-1} \over \ln(1+\ctwo)-\ctwo/(1+\ctwo)}
\ ,
\label{e:nfw}
\end{equation} 
\item the Hernquist model \citep{Hernquist90},
\begin{equation}
M(r)={\mtwo \, (\rh+\rtwo)^2 \over \rtwo^2} {r^2 \over (r+\rh)^2}
\ ,
\label{e:her}
\end{equation} 
where $\rh=2 \, \rs$,
\item the Einasto model \citep{Einasto65,MBM10,Tamm+12},
\begin{equation}
M(r)=\mtwo {P[3m,2m \, (r/\rs)^{1/m}] \over P[3m,2m \, (\rtwo/\rs)^{1/m}]} \,
\end{equation}
where $P(a,x)=\gamma(a,x)/\Gamma(a)$ is the regularized incomplete
gamma function, and where we fix $m=5$, a typical value for
cluster-size halos extracted from cosmological numerical simulations
\citep{MBM10},
\item the Burkert model \citep{Burkert95},
\begin{eqnarray}
M(r) = \mtwo \, \{ \ln [1+(r/\rb)^2] + 2 \ln (1+r/\rb) \nonumber \\
- 2 \arctan (r/\rb) \} \times \{\ln [1+(\rtwo/\rb)^2] \nonumber \\
+ 2 \ln (1+\rtwo/\rb) - 2 \arctan (\rtwo/\rb) \}^{-1} \ ,
\label{e:bur}
\end{eqnarray} 
where $\rb \simeq 2/3 \, \rs$,
\item the Softened Isothermal Sphere \citep[SIS model, hereafter; see e.g.][]{GDK99}
\begin{equation}
M(r) = \mtwo {r/\ri-\arctan(r/\ri) \over \rtwo/\ri-\arctan (\rtwo/\ri)}
\ ,
\label{e:sis}
\end{equation} 
where $\ri$ is the core radius.
\end{enumerate}
The NFW and Hernquist mass density profiles are characterized by
central logarithmic slopes $\gamma=-1$, while the Burkert and SIS
mass density profiles have a central core, $\gamma=0$. Somewhat in
  between these two extremes, the Einasto profile has not a fixed
  central slope but one that asymptotically approaches zero near the
  center, $\gamma=-2 \, (r/\rs)^{1/m}$. The asymptotic slopes of the
NFW, Hernquist, Burkert, and SIS mass density profiles are $\gamma=-3,
-4, -3,$ and $-2$, respectively.  The NFW and the Einasto models
have been shown to successfully describe the mass density profiles of
observed clusters (see Section~\ref{s:intro}). The Hernquist model is
well studied \citep[e.g.][]{BD02} and it has been shown to provide a
good fit to the mass profile of galaxy clusters
\citep{Rines+00,Rines+01,RGKD03,RD06}. This is also true of the
Burkert model \citep{KBM04,BS06}, but not of the SIS
model \citep{RGKD03,KBM04}.

As for $\br$, we consider the following models:
\begin{enumerate}
\item 'C': Constant anisotropy with radius, $\beta=\beta_C$;
\item 'T': from \citet{Tiret+07},
\begin{equation}
\beta_{\rm T}(r)=\beta_{\infty} \, {r \over r+\rs} \ ,
\label{e:T}
\end{equation}
isotropic at the center, with
anisotropy radius identical to $\rs$, characterized by the
anisotropy value at large radii, $\beta_{\infty}$;
\item 'O': anisotropy of opposite sign at the center and at large radii,
\begin{equation}
\beta_{\rm O}(r)=\beta_{\infty} \, {r-\rs \over r+\rs} \ ,
\label{e:O}
\end{equation}
\end{enumerate}
The C model is the simplest, and has been frequently used in previous
studies \citep[e.g.][]{Merritt87,vanderMarel+00,LM03}.  The T model
has been shown by \citet{MBM10,MBB13} to provide a good fit to the
velocity anisotropy profiles of cosmological cluster-mass halos. Here
we introduce the O model to account for the possibility of deviation
from the general behavior observed in numerically simulated halos --
the O model allows for non-isotropic orbits near the cluster center
while the T model does not.  Isotropic orbits are allowed in all three
models. Note that the $\rs$ parameter common to the T and O models is
the same parameter that enters the NFW and Einasto $M(r)$ models,
and is related to the scale parameters of the Hernquist and Burkert
$M(r)$ models. For the SIS model $\rs$ cannot be uniquely defined,
hence we can only consider the C $\br$ model, and not the T and O
models.

\begin{table*}
\centering
\caption{Results of the MAMPOSSt analysis.}
\label{t:mamposst}
\begin{tabular}{ll|cccc|cccc}
\hline 
\multicolumn{2}{c|}{Models} & $\rtwo$ & $\rr$ & Vel. & Lik. & $\rtwo$ & $\rr$ & Vel. & Lik. \\
$M(r)$ & $\br$ & [Mpc]   & [Mpc] & anis. & ratio & [Mpc]   & [Mpc] & anis. & ratio \\
& & & & & & & & \\
\hline
& & \multicolumn{4}{c|}{$\rn=0.74$ Mpc}  & \multicolumn{4}{c}{$\rn=0.63$ Mpc} \\
\hline
& & & & & & & & \\
NFW & C & $1.97_{-0.12}^{+0.06}$ & $0.43_{-0.06}^{+0.78}$ & $0.4_{-0.1}^{+0.3}$ &
 1.00  & $1.99_{-0.09}^{+0.08}$ & $0.39_{-0.06}^{+0.65}$ & $0.4_{-0.1}^{+0.3}$ &
 1.00 \\
& & & & & & & & \\
NFW & T & $1.94_{-0.13}^{+0.05}$ & $0.36_{-0.02}^{+0.33}$ & $0.5_{-0.0}^{+0.4}$ &
 0.87 & $1.96_{-0.11}^{+0.05}$ & $0.34_{-0.02}^{+0.31}$ & $0.5_{-0.0}^{+0.4}$ &
 0.88 \\
& & &  & & & & & \\
NFW & O & $1.94_{-0.10}^{+0.07}$ & $0.28_{-0.04}^{+0.15}$ & $0.5_{-0.2}^{+0.4}$ &
 0.62  & $1.96_{-0.10}^{+0.07}$ & $0.27_{-0.04}^{+0.14}$ & $0.5_{-0.2}^{+0.4}$ &
 0.65 \\
& & & & & & & & \\
Hernquist & C & $2.00_{-0.13}^{+0.06}$ & $1.13_{-0.13}^{+1.56}$ & $0.5_{-0.1}^{+0.3}$ &
 0.89  & $2.03_{-0.10}^{+0.07}$ & $1.07_{-0.15}^{+1.28}$ & $0.5_{-0.1}^{+0.3}$ &
 0.88 \\
& & & & & & & & \\
Hernquist & T & $1.97_{-0.11}^{+0.05}$ & $0.97_{-0.06}^{+0.59}$ & $0.6_{-0.0}^{+0.3}$ &
 0.64  & $2.00_{-0.10}^{+0.06}$ & $0.92_{-0.06}^{+0.56}$ & $0.6_{-0.0}^{+0.3}$ &
 0.64 \\
& & & & & & & & \\
Hernquist & O & $1.98_{-0.09}^{+0.07}$ & $0.72_{-0.10}^{+0.28}$ & $0.4_{-0.2}^{+0.5}$ &
 0.34  & $1.99_{-0.09}^{+0.07}$ & $0.70_{-0.09}^{+0.27}$ & $0.4_{-0.2}^{+0.5}$ &
 0.35 \\
& & & & & & & & \\
Einasto & C & $1.98_{-0.14}^{+0.06}$ & $0.47_{-0.05}^{+0.88}$ & $0.4_{-0.1}^{+0.3}$ &
 1.00  & $2.01_{-0.11}^{+0.07}$ & $0.42_{-0.05}^{+0.74}$ & $0.4_{-0.1}^{+0.3}$ &
 1.00 \\
& & & & & & & & \\
Einasto & T & $1.95_{-0.13}^{+0.04}$ & $0.41_{-0.02}^{+0.34}$ & $0.6_{-0.0}^{+0.4}$ &
 0.86  & $1.98_{-0.12}^{+0.05}$ & $0.39_{-0.02}^{+0.33}$ & $0.6_{-0.0}^{+0.4}$ &
 0.87 \\
& & & & & & & & \\
Einasto & O & $1.95_{-0.10}^{+0.07}$ & $0.31_{-0.04}^{+0.16}$ & $0.5_{-0.2}^{+0.4}$ &
 0.57  & $1.98_{-0.10}^{+0.07}$ & $0.30_{-0.04}^{+0.15}$ & $0.5_{-0.2}^{+0.4}$ &
 0.59 \\
& & & & & & & & \\
Burkert & C & $1.99_{-0.09}^{+0.08}$ & $0.30_{-0.06}^{+0.33}$ & $0.5_{-0.1}^{+0.3}$ &
 0.74  & $2.01_{-0.07}^{+0.09}$ & $0.27_{-0.05}^{+0.28}$ & $0.5_{-0.1}^{+0.3}$ &
 0.73 \\
& & & & & & & & \\
Burkert & T & $1.96_{-0.10}^{+0.05}$ & $0.23_{-0.02}^{+0.16}$ & $0.5_{-0.0}^{+0.4}$ &
 0.51  & $1.98_{-0.08}^{+0.06}$ & $0.22_{-0.02}^{+0.15}$ & $0.5_{-0.0}^{+0.4}$ &
 0.52 \\
& & & & & & & & \\
Burkert & O & $1.96_{-0.09}^{+0.07}$ & $0.18_{-0.03}^{+0.07}$ & $0.4_{-0.2}^{+0.5}$ &
 0.33  & $1.97_{-0.09}^{+0.07}$ & $0.17_{-0.03}^{+0.07}$ & $0.4_{-0.2}^{+0.5}$ &
 0.34 \\
& & & & & & & & \\
SIS & C & $1.83_{-0.09}^{+0.10}$ & $0.01_{-0.00}^{+0.02}$ & $0.5_{-0.2}^{+0.3}$ &
 0.44  & $1.88_{-0.10}^{+0.09}$ & $0.01_{-0.00}^{+0.03}$ & $0.5_{-0.2}^{+0.3}$ &
 0.37 \\
& & & & & & & & \\
\hline 
\end{tabular}
\tablefoot{Results of the MAMPOSSt analysis are shown as obtained
    using two different input values of the best-fit $\rn$ parameter,
    determined outside MAMPOSSt (see Sect.~\ref{ss:compl} and
  Table~\ref{t:slos}).  1 $\sigma$ marginalized errors are listed for
  all free parameters in the MAMPOSSt analysis.  The scale radius
  $\rr$ is $\rs$ for the NFW and Einasto models, $\rh,\rb,$ and
  $\ri$, for the Hernquist, Burkert, and SIS $M(r)$ models,
  respectively.  The velocity anisotropy ('Vel. anis.') is $\beta_C$
  for the C model and $\beta_{\infty}$ for the T and O models. The
  likelihood ('Lik.')  ratios are given relative to the maximum among
  the models.}
\end{table*}

\begin{figure}
\begin{center}
\begin{minipage}{0.5\textwidth}
\resizebox{\hsize}{!}{\includegraphics{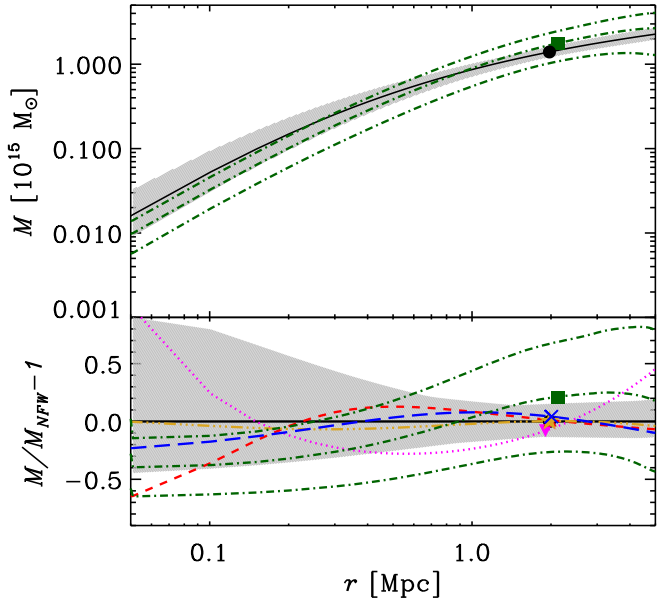}}
\end{minipage}
\end{center}
\caption{{\em Top panel:} Mass profiles as obtained from the
    MAMPOSSt and Caustic analyses. The MAMPOSSt result is that
    obtained using the NFW model and the O $\br$ model, and is
    represented by a black curve within a grey shaded area (1 $\sigma$
    confidence region). The Caustic result is represented by green
    dash-dotted curves (central value within 1 $\sigma$ confidence
    region). The black dot and green square represent the locations of
    the $[\rtwo,\mtwo]$ values for the MAMPOSSt and Caustic $M(r)$.
  {\em Bottom panel:} Fractional difference between different mass
    profiles and the MAMPOSSt best-fit to the NFW $M(r)$ with O $\br$
    model (displayed in the top panel).  The MAMPOSSt best-fit O $\br$
    Hernquist, Einasto, and Burkert models are represented by the blue
    long-dashed, gold triple-dot-dashed, and red short-dashed curves,
    respectively. The MAMPOSSt best-fit C $\br$ SIS model is
    represented by the magenta dotted curve. The Caustic $M(r)$ and 1
    $\sigma$ confidence levels are represented by the green
    dash-dotted curves. The solid line marks the zero and the grey
    shaded area the 1 $\sigma$ confidence region of the NFW model fit.
    Symbols represent the location of the
    $[\rtwo,\mtwo/M_{NFW}(\rtwo)-1]$ values for the different mass
    profiles, NFW (filled black dot), Hernquist (blue X), Einasto
    (gold star), Burkert (red triangle), SIS (magenta inverted
    triangle), Caustic (green square). The NFW and Burkert values are
    barely visible in the plot, because they are virtually
    indistinguishable from the Einasto and Hernquist values. All
    MAMPOSSt results displayed here are for the $\rn=0.63$ Mpc value
    (see Table~\ref{t:mamposst}).}
\label{f:mr}
\end{figure}

\begin{figure}
\begin{center}
\begin{minipage}{0.5\textwidth}
\resizebox{\hsize}{!}{\includegraphics{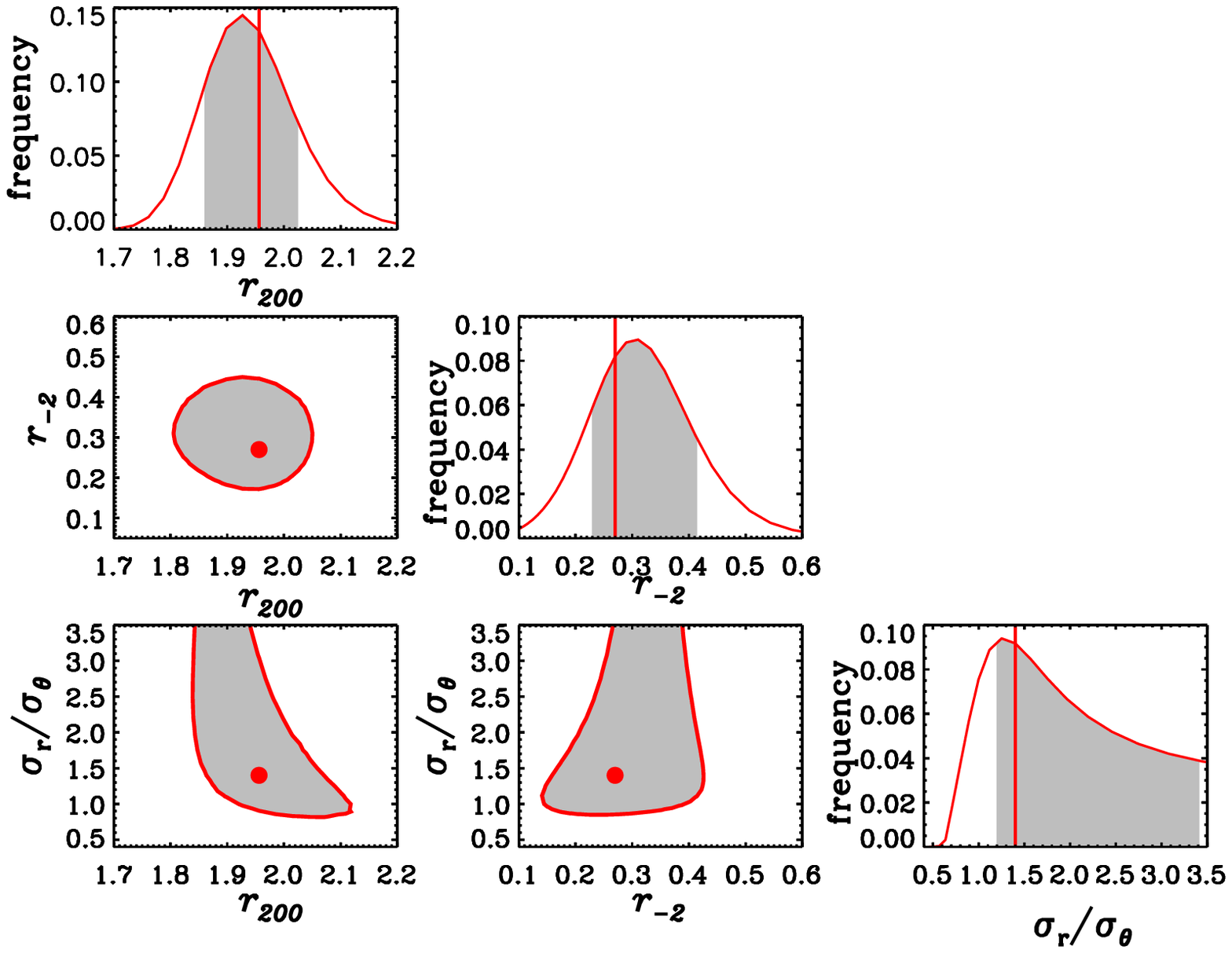}}
\end{minipage}
\end{center}
\caption{Results of the MAMPOSSt analysis using the NFW and O models
  for $M(r)$ and $\br$, respectively. The vertical lines and dots
  indicate the best-fit solutions. The likelihood distributions on
  each parameter are obtained by marginalizing vs. the other two
  parameters. Gray-shading in the likelihood distribution plots
  indicate the 1 $\sigma$ confidence regions. The red, gray-shaded
  contours are 1 $\sigma$ confidence levels on the two labeled
  parameters, obtained by marginalizing vs. the third parameter. Note
  that we show results for $\sigma_r/\sigma_\theta$ rather than for
  $\beta$ (see eq.~\ref{e:beta}).}
\label{f:mam_con}
\end{figure}

In total, we run MAMPOSSt with 3 free parameters, i.e. the virial
radius $\rtwo$, the scale radius of the total mass distribution $\rr$
(equal to $\rs, \rh, \rb$ or $\ri$, depending on the $M(r)$ model),
and the anisotropy parameter, $\beta_C$ or $\beta_{\infty}$.  Note
that we do {\em not} assume that light traces mass, i.e. we allow the
scale radius of the total mass distribution to be different from that
of the galaxy distribution, $\rr \neq \rn$.  The results of the
MAMPOSSt analysis are given in Table~\ref{t:mamposst}. The best-fits
are obtained using the NEWUOA software for unconstrained optimization
\citep{Powell06}. The errors on each of the parameters listed in the
table are obtained by a marginalization procedure, i.e. by integrating
the probabilities $p(\rtwo,\rr,\beta)$ provided by MAMPOSSt, over the
remaining two free parameters. 

In Table~\ref{t:mamposst} we list two sets of results, one for
each of the best-fit values of $\rn$ found in Sect.~\ref{ss:compl}. 
The results are very similar in the two cases. On average, the values
of $\rtwo, \rr,$ and $\beta$ or $\beta_{\infty}$ change by 2, 5, and 2~\%,
respectively. These variations are much smaller than the statistical
errors on the parameters, therefore we only consider the set of results
obtained for $\rn=0.63$ Mpc, in the following (this is the value obtained
for the complete sample of spectroscopic + photometric cluster members,
see Sect.~\ref{ss:compl}).

Using the likelihood-ratio test \citep{Meyer+75} we find that all
models are statistically acceptable (likelihood ratios are listed in
the last column of Table~\ref{t:mamposst}). This is also visible from
Fig.~\ref{f:mr} where we display the five $M(r)$ corresponding to the
best-fit NFW, Hernquist, Einasto, and Burkert models with O
$\br$, and to the best-fit SIS model with $C$ $\br$.  The SIS $M(r)$
is in some tension with the others due to the fact that it is
essentially a single power-law, as the value of its core radius $\ri$
is constrained by the MAMPOSSt analysis to be very small (see Table
~\ref{t:mamposst}).

The different models give best-fit values of $\rtwo$ in agreement
within their 1~$\sigma$ errors. The rms of all $\rtwo$ values is 0.04,
smaller than the error on any individual $\rtwo$ value. This is
also true of the $\rs$ parameter (we use the appropriate scaling
factors to convert $\rh$ and $\rb$ to $\rs$), for which the rms is
0.08, and of the anisotropy parameter for which the rms is 0.06.

Since the uncertainties on the values of the parameters are dominated
by statistical errors, and not by the systematics induced by the model
choice, for simplicity in the rest of this paper we only consider the
MAMPOSSt results obtained for one of the considered models. We choose
the NFW model for $M(r)$, for the sake of comparing our results to
those of \citetalias{Umetsu+12}, and also because it provides slightly
higher likelihoods than the Hernquist, Burkert, and SIS mass
models (for fixed $\br$ model) and comparable likelihoods to those
  of the Einasto model. As for the $\br$ model, we choose the O
model, since it is the one that gives the smallest errors on the
$M(r)$ parameters, in the sense of maximizing the figure of merit
FoM$\, \equiv (\rtwo \, \rs)/(\delta \rtwo \, \delta \rs)$, where
$\delta \rtwo$ and $\delta \rs$ are the (symmetrized) errors on,
respectively, $\rtwo$ and $\rs$. In Fig.~\ref{f:mam_con} we display
the results of the MAMPOSSt analysis for the NFW+O models.  In
Sect.~\ref{s:beta} we will show how the best-fit $\br$ models for the
NFW mass model compare with our non-parametric $\br$ determination
from the Jeans inversion (see Fig.~\ref{f:beta}).

\subsection{The Caustic method}
\label{ss:CAU}
The Caustic method \citep{DG97,Diaferio99} is based on the
identification of density discontinuities in the $R, \vrf$ space. This
method does not require the assumption of dynamical equilibrium
outside the virial region, hence it makes use of all galaxies, not
only of member galaxies, and can provide $M(r)$ also at $r> \rtwo$.
Moreover the method does not require to assume a model for
$M(r)$. This comes at the price of some simplifying assumptions that
can induce systematic errors, as we see below.

\begin{figure}
\begin{center}
\begin{minipage}{0.5\textwidth}
\resizebox{\hsize}{!}{\includegraphics{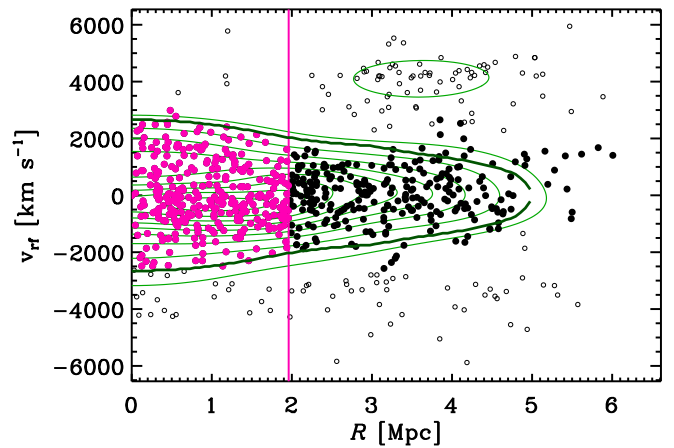}}
\end{minipage}
\end{center}
\caption{Caustics in the $R, \vrf$ space; the thick-line
  caustic is the one identified following the prescription of
  \citet{Diaferio99}.  Filled dots identify members selected using the
  P+G method; the vertical line indicates the location of $\rume$.}
\label{f:cau}
\end{figure}

\begin{figure}
\begin{center}
\begin{minipage}{0.5\textwidth}
\resizebox{\hsize}{!}{\includegraphics{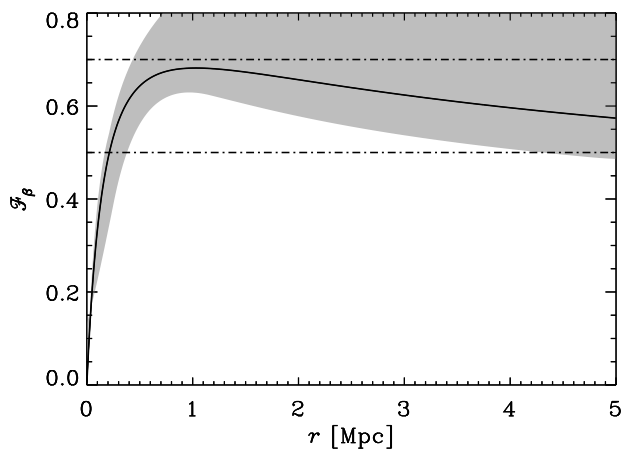}}
\end{minipage}
\end{center}
\caption{The ${\cal F}_{\beta}$ function obtained by adopting the
  best-fit $M(r)$ NFW model with an O $\br$ model, using MAMPOSSt
  (solid black curve) within 1 $\sigma$ confidence region (hatched
  gray region).  The two horizontal dashed lines indicate two commonly
  adopted constant values of ${\cal F}_{\beta}$ in the literature.}
\label{f:fbeta}
\end{figure}

In Fig.~\ref{f:cau} we show the projected phase-space distribution of
all galaxies and galaxy iso-number density contours, computed using
Gaussian adaptive kernels with an initial 'optimal' kernel size
\citep[as defined in][]{Silverman86}. Before estimating the density
contours, rest-frame velocities and clustercentric distances are
scaled in such a way as to have the same dispersion for the scaled
radii and scaled velocities. The data-set is mirrored across the $R=0$
axis before the density contours are estimated, to avoid edge-effects
problems. To choose the density threshold that defines the contour
(the 'caustic') to use, we follow the prescriptions of
\citet{Diaferio99}, which depend on an estimate of the velocity
dispersion of cluster members. We use the P+G cluster membership
definition (Section~\ref{ss:members}), for consistency with the rest
of our dynamical analyses in this paper.

\begin{table*}
\centering
\caption{Best-fit dynamical parameters for the NFW $M(r)$ model.}
\label{t:NFW}
\begin{tabular}{lllllll}
\hline 
Method & Sample & $N_{members}$ & $\rtwo$ & $\rs$ & $\mtwo$ & $\ctwo$ \\
                  &                    & [Mpc]      & [Mpc] &    [$10^{15} \, \msun$]     &      \\
\hline  
& & & & \\
$\slos$+$\rn$ & $R \leq 1.98$ Mpc (passive only) & 261 & {\em 1.98} $\pm$ {\em 0.10}   &  {\em 0.63}$_{-0.09}^{+0.11}$ & {\em 1.41} $\pm$ {\em 0.21} & {\em 3.1} $\pm$ {\em 0.5} \\
& & & & \\
MAMPOSSt    & $0.05 \leq R \leq 1.96$ Mpc  & 330 & $1.96_{-0.10}^{+0.07}$ & $0.27_{-0.04}^{+0.14}$ & $1.37 \pm 0.18$ & $7.3 \pm 2.4$ \\
& & & & \\
Caustic      & $R \leq 2 \times 1.96$ Mpc & 527 & $2.08_{-0.30}^{+0.09}$ & $0.47_{-0.09}^{+0.47}$ & $1.63 \pm 0.58$ & $4.4 \pm 3.0$ \\
& & & & \\
MAMPOSSt+Caustic &  &  & $1.96_{-0.09}^{+0.14}$ & $0.35_{-0.09}^{+0.14}$ & $1.37 \pm 0.24$ & $5.6 \pm 1.9$ \\
& & & & \\
Lensing   & \citetalias{Umetsu+12} &  & $1.96 \pm 0.11$ & $0.34 \pm 0.06$ &  $1.37 \pm 0.23$ &  $5.8 \pm 1.1$ \\
& & & & \\
\hline
\end{tabular}
\tablefoot{$N_{members}$ is the number of cluster members in the
    different samples used for the dynamical analyses. The results of
  the $\slos$ + $\rn$ method are listed in italic to indicate that
  they are based on the simplified assumptions that light traces mass
  and that the galaxy and DM particle velocity distributions are
  identical. These assumption are dropped for the MAMPOSSt and Caustic
  methods. The error on $\rtwo$ and that on $\rs$ are obtained by
  marginalizing on the other parameters.  The errors on $\mtwo$ and
  $\ctwo$ are derived from propagating the symmetrized errors on
  $\rtwo$ and $\rs$.  The line labelled
  'MAMPOSSt+Caustic' lists the results obtained by the combination
    of the MAMPOSSt and Caustic solutions. These results are therefore
    based on the samples used separately for the MAMPOSSt and Caustic
    methods. Since the two samples largely overlap and
    the two methods are not entirely independent, the errors 
    are in this case multiplied by
  $\sqrt{2}$.}
\end{table*}

The velocity amplitude of the chosen caustic is related to $M(r)$ via
a function of both the gravitational potential and $\br$, called
${\cal F}_{\beta}$.  For simplicity most studies \citep[with the
  notable exception of][]{BG03} have so far used constant ${\cal
  F}_{\beta}(r)$, following the initial suggestion of \citet{DG97} and
\citet{Diaferio99}. With the most recent implementation of the caustic
algorithm by \citet{Serra+11}, the value of ${\cal F}_{\beta}=0.7$ was
adopted. The value ${\cal F}_{\beta}=0.5$ preferred by \citet{DG97},
\citet{Diaferio99}, and \citet{GDRS13} was appropriate for an earlier
implementation of the algorithm that however tended to overestimate
the escape velocity by 15--20\% on average.  

The unknown value of ${\cal F}_{\beta}$ is a major systematic
uncertainty in this method.  The correct value of ${\cal F}_{\beta}$
to use might be different for different membership definitions, as
suggested by the analysis of numerically simulated halos of
\citet{Serra+11}.  For consistency we use for the Caustic method the
same membership definition used for the MAMPOSSt analysis (see
Sect.~\ref{ss:MAM}).  We can therefore take advantage of our
MAMPOSSt-based determinations of $M(r)$ and $\br$ to determine ${\cal
  F}_{\beta}$ for the Caustic method.

We adopt the best-fit NFW $M(r)$ + O $\br$ model (see
Table~\ref{t:mamposst}) and obtain the ${\cal F}_{\beta}(r)$ shown in
Fig.~\ref{f:fbeta}. The large uncertainty associated to the
$\beta_{\infty}$ parameter of the O model propagates into a large
uncertainty on ${\cal F}_{\beta}$. Within the uncertainties ${\cal
  F}_{\beta}(r)$ is consistent with the value of 0.7 but only at
radii $r>0.5$ Mpc. It is instead inconsistent with the value of
  0.5 at most radii. Over most of the radial range, ${\cal
  F}_{\beta}(r)$ is intermediate between these two commonly adopted
constant values, but not near the center, where it is smaller.
Constant-${\cal F}_{\beta}$ Caustic determinations of $M(r)$ are known
to suffer from an overestimate at small radii \citep{Serra+11}; the
radial dependence of our adopted ${\cal F}_{\beta}(r)$ is likely to
correct for this bias.

The uncertainties in the Caustic $M(r)$ estimate are derived following
the prescriptions of \citet{Diaferio99}.  According to
\citet{Serra+11} these prescriptions lead to estimate 50\% confidence
levels; we therefore multiply them by 1.4 to have $\sim 1
\sigma$ confidence levels. 

The Caustic $M(r)$ within its 1 $\sigma$ confidence region is shown in
Fig.~\ref{f:mr}. It is consistent with the $M(r)$ obtained via the
MAMPOSSt method. This consistency is at least partly
  enforced by the fact that we calibrated ${\cal F}_{\beta}(r)$ using
  the results we obtained with MAMPOSSt.

We obtain the mass density profile
$\rho(r)$ from numerical differentiation of the Caustic $M(r)$, and
then fit the NFW model, limiting the fit to radii below twice
$\rume$ (we can extend the
  fit beyond $\rume$ because the Caustic method is not based on the
  assumption of dynamical equilibrium).  The best-fit is obtained
from a $\chi^2$-minimization procedure. Uncertainties in the best-fit
value are obtained using the $\chi^2$ distribution, by setting the
effective number of independent data to the ratio between the used
radial range in the fit and the adaptive-kernel radial scale used to
determine the caustic itself.  The NFW model provides a good fit to
the Caustic $\rho(r)$ over the full radial range considered (reduced
$\chi^2=0.4$).

The best-fit $\rtwo$ and $\rs$ values of the NFW model fitted
  to the Caustic-derived mass density
  profile, and their marginalized 1 $\sigma$ errors,
are listed in Table~\ref{t:NFW}. For comparison, we also list in the
same Table the adopted results of the MAMPOSSt analysis
(Sect.~\ref{ss:MAM}). The MAMPOSSt and Caustic values of $\rtwo$ and
$\rs$ are consistent within their error bars.

\subsection{Combining different mass profile determinations}
\label{ss:combMr}
\begin{figure}
\begin{center}
\begin{minipage}{0.5\textwidth}
\resizebox{\hsize}{!}{\includegraphics{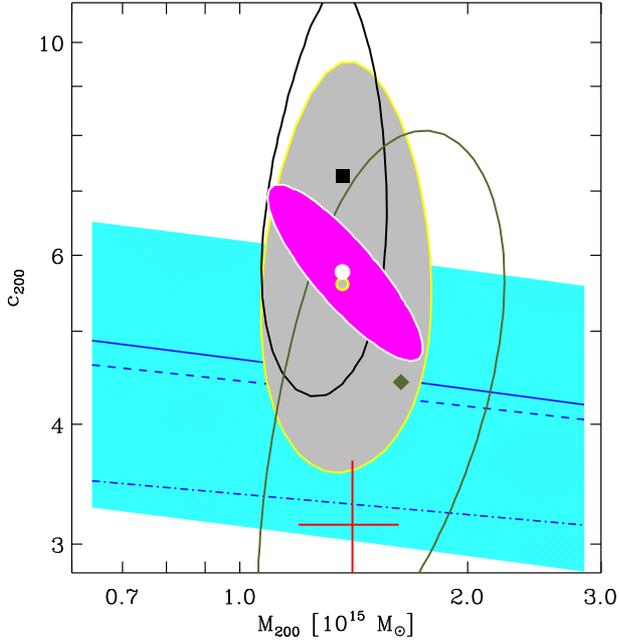}}
\end{minipage}
\end{center}
\caption{Best-fit solutions and 1 $\sigma$ contours in the
  $\mtwo$-$\ctwo$ space for the NFW $M(r)$ model 
    (see also Table~\ref{t:NFW}).  Lensing
  analysis of \citetalias{Umetsu+12}: small magenta-filled region
  (with white border) and white filled dot.  MAMPOSSt analysis: black
  vertically-elongated contour and filled square.  Caustic analysis:
  green horizontally-elongated contour and green diamond. Joint
  MAMPOSSt + Caustic constraints: gray-filled region and gray dot with
  yellow borders. Best-fit value and 1 $\sigma$ error bars from
    the $\slos$+$\rn$ analysis: big
  red cross. The solid (resp. dashed) blue line and shaded cyan region
  represent the theoretical $cMr$ of \citet{BHHV13} for relaxed
  (resp. all) halos and its 1 $\sigma$ scatter. The dash-dotted blue
  line represents the theoretical $cMr$ of \citet{DeBoni+13} for
  relaxed halos.}
\label{f:cMr}
\end{figure}

\begin{figure}
\begin{center}
\begin{minipage}{0.5\textwidth}
\resizebox{\hsize}{!}{\includegraphics{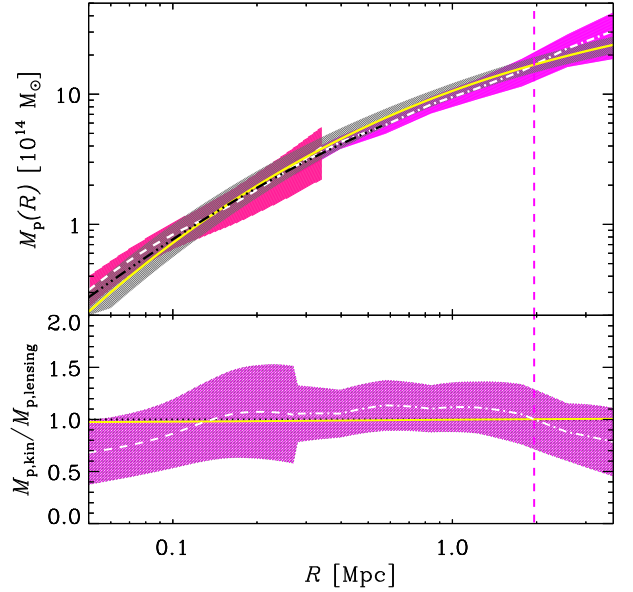}}
\end{minipage}
\end{center}
\caption{{\it Top panel:} The projected mass profile $M_{\rm p}(R)$
  from the joint MAMPOSSt+Caustic pNFW solution (solid yellow line)
  within 1~$\sigma$ confidence region (hatched gray region), and from
  the lensing analysis of \citetalias{Umetsu+12} (dashed white line:
  strong lensing analysis; dash-dotted line: weak lensing analysis,
  after subtraction of the contribution of the large-scale structure
  along the line-of-sight) within 1~$\sigma$ confidence region
  (hatched magenta regions).  The black triple-dots-dashed line is the
  pNFW mass profile from \citetalias{Umetsu+12}'s analysis of Chandra
  data.  The vertical dashed line indicates the location of $\rume$ in
  both panels.  {\it Bottom panel: } The ratio between the kinematic
  and lensing determinations of $M_{\rm p}(R)$. The white dashed and
  dash-dotted (resp. solid yellow) line represents the ratio obtained
  using the non-parametric determination (resp. the pNFW
  parametrization) of the lensing $M_{\rm p}(R)$. The pink hatched
  region represents the confidence region of this ratio for the
  non-parametric $M_{\rm p}(R)$ lensing solution. The horizontal black
  dotted line indicates the value of unity.}
\label{f:mproj}
\end{figure}

\begin{figure}
\begin{center}
\begin{minipage}{0.5\textwidth}
\resizebox{\hsize}{!}{\includegraphics{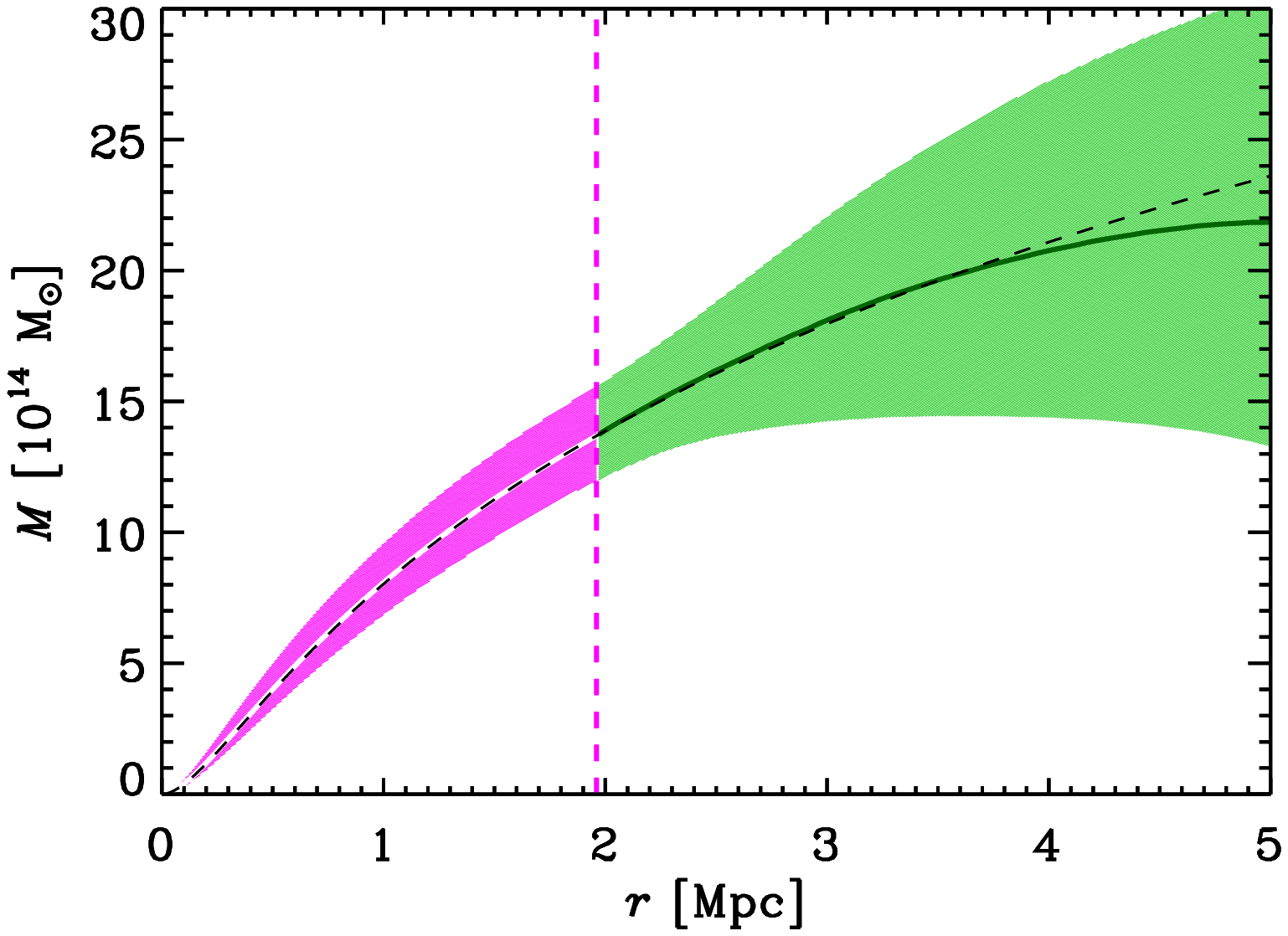}}
\end{minipage}
\end{center}
\caption{The solid (white and green) curve and hatched (magenta and
  green) region represent our fiducial $M(r)$ within 1~$\sigma$
  confidence levels. This corresponds to the NFW best-fit to the
  lensing mass profile of \citetalias{Umetsu+12} (white curve within
  magenta region) out to $\rume$ (indicated by a vertical
  dashed line), and to the Caustic non-parametric mass profile (green
  curve within light green region) beyond that radius.  The dashed
  black curve represents the NFW best-fit solution obtained by the
  combined MAMPOSSt+Caustic analysis.}
\label{f:massprof}
\end{figure}

In Sections~\ref{ss:MAM} and \ref{ss:CAU} we have found that the NFW
model is the best description of $M(r)$ among the three we have
considered. This is a particularly welcome result because also
\citetalias{Umetsu+12} found that the NFW model is a good description
to the cluster $M(r)$ obtained by a gravitational lensing analysis. It
is therefore straightforward to compare our results with those of
\citetalias{Umetsu+12}. 

In Table~\ref{t:NFW} we list the values of $\rtwo, \rs$ and of the
related parameters $\mtwo, \ctwo$ of the NFW model, as obtained from
the MAMPOSSt and Caustic analyses (see Sect.~\ref{ss:MAM} and
\ref{ss:CAU}), as well as the results obtained by
\citetalias{Umetsu+12}. In addition, we list the values obtained by
combining the constraints from the MAMPOSSt and Caustic analyses. The
combination is done by summing the $-2 \ln L$ values from the MAMPOSSt
analysis (where $L$ are the likelihood values) and the $\Delta \chi^2$
values from the NFW fit to the Caustic mass density profile, and by
taking the value corresponding to the mimimum sum. To account for the
fact that the two methods are largely based on the same data-set, the
marginalized errors on the resulting parameters are multiplied by
$\sqrt{2}$. Combining the MAMPOSSt and Caustic results allows us to
reach an accuracy on the $M(r)$ parameters which is unprecedented for
a kinematic analysis of an individual cluster, similar to that
obtained from the combined strong and weak lensing analysis. There is
a very good agreement between the $\rtwo,\rs$ values obtained by the
combined MAMPOSSt and Caustic analyses and those obtained by the
lensing analysis of \citetalias{Umetsu+12}.

Our kinematic constraints on the cluster $M(r)$ are free of the usual
assumptions that light traces mass and that the DM particle and galaxy
velocity distributions are identical.  When dealing with poor
data-sets (unlike the one presented here) one is forced to adopt
simpler techniques and accept these assumptions. It is instructive to
see what we would obtain in this case. We would use the sample of
passive members to infer the value of $\rtwo$ from the $\slos$ value,
as we have done in Sect.~\ref{ss:members}. As for the value of $\rs$
we would assume it to be identical to $\rn$ (see Sect.~\ref{ss:compl});
this is the so-called 'light traces mass' hypothesis.  There is some
observational support that this assumption is verified (on average)
for the passive population of cluster members
\citep[e.g.][]{vanderMarel+00,BG03,KBM04}.  In Table~\ref{t:NFW} we
list the $\slos$-based value of $\rtwo$, the $\rn$ value of the
spatial distribution of passive cluster members, and the implied
values of $\mtwo,\ctwo$ (we label the method '$\slos$+$\rn$'
hereafter).  Formally the statistical uncertainties on these values
are smaller than those of any other method.  However, this comes at a
price of biasing the inferred value of $\ctwo$ low, since the
'light traces mass' hypothesis does not seem to be verified in this
cluster, i.e. $\rr \neq \rn$. On the other hand, the $\mtwo$ value is
in excellent agreement with those derived using more sophisticated
methods.

In Fig.~\ref{f:cMr} we show the best-fit solutions and 1~$\sigma$
contours for the NFW $M(r)$ parameters $\mtwo,\ctwo$, as obtained with
the MAMPOSSt and Caustic analyses, as well as the results obtained by
\citetalias{Umetsu+12}. Interestingly, the covariance between the
  errors in the $\mtwo$ and $\ctwo$ parameters is different for the
  different techniques (MAMPOSSt, Caustic, and lensing).  We also
show the results obtained from the simplified
$\slos$+$\rn$ method and the results from the combined MAMPOSSt
  and Caustic solution, where we take care of drawing the contours at
  a level twice as high as that used for the individual MAMPOSSt and
  Caustic solutions.

In Fig.~\ref{f:cMr} we also show theoretical predictions for the
$cMr$ of the total halo mass distribution. From the DM-only simulations
of \citet{BHHV13} we show two $cMr$, one for all halos in their
cosmological simulations, and another for the subset of dynamically
relaxed halos. From the hydrodynamical simulations of
\citet{DeBoni+13} we only show the $cMr$ for relaxed halos.  Our
$\mtwo,\ctwo$ results are in reasonable agreement with theoretical
predictions.  The difference between the observed and predicted
$\mtwo,\ctwo$ values is smaller than both the observational
uncertainties and the theoretical scatter in the $cMr$. Our result is in
better agreement with the theoretical prediction from the DM-only
simulations of \citet{BHHV13} than with that from the hydrodynamic
simulation of \citet{DeBoni+13}. Our result lies at the high
concentration end of the allowed theoretical range, a region occupied
by more dynamically relaxed halos in numerical simulations
\citep[e.g.][]{Maccio+07,DeBoni+13,BHHV13}. This is consistent with
the fact that this cluster was selected to be free of signs of ongoing
mergers \citep{Postman+12}. Also the good agreement between the
lensing, and the kinematic estimates of the cluster mass profile is an
indication for dynamical relaxation. Deviation from relaxation should
in fact affect the kinematic analysis but not the lensing analysis,
and we should not obtain consistent results from the two analyses.

Independent constraints on the cluster $M(r)$ have also been obtained
from the analysis of Chandra X-ray data by \citetalias{Umetsu+12}. The
X-ray data do not allow estimating $M(r)$ beyond $\rfive$. We can
however directly compare the $M(r)$ obtained by the different methods
in the radial range where they overlap.  Since the lensing technique
provides the {\em projected} $M(r)$, $M_{\rm p}(R)$, for the sake of
comparison we also project the NFW models that provide the best-fit to
the kinematic and X-ray data. In Fig.~\ref{f:mproj} we show
\citetalias{Umetsu+12}'s strong and weak lensing
determinations\footnote{The weak lensing solution we display here is
  the one obtained by \citetalias{Umetsu+12} after removal of an
  extended large-scale structure feature contaminating the external
  regions of the cluster along the line-of-sight. See
  \citetalias{Umetsu+12} for details.} of $M_{\rm p}(R)$, within their
1~$\sigma$ confidence regions, as well as the pNFW model best-fit
obtained by \citetalias{Umetsu+12} using Chandra X-ray data, and the
pNFW model best-fit we obtained by the joint MAMPOSSt+Caustic
likelihood analysis. The agreement between the different mass profile
determinations is very good\footnote{Note that in this Figure we show
  the non-parametric solution for $M_{\rm p}(R)$ obtained by the
  lensing technique, not the pNFW fit.}.

Given the good consistency between the $M(r)$ parameter values
obtained by the kinematic and lensing techniques, we now combine them
to form a unique $M(r)$ solution. Within $\rume$ we adopt the best-fit
NFW $M(r)$ obtained by the lensing analysis of \citetalias{Umetsu+12},
since this has the smallest uncertainties, as measured by the figure
of merit defined in Sect.~\ref{ss:MAM}.  Beyond $\rume$ we adopt the
$M(r)$ determination obtained by the Caustic technique. In fact, the
lensing analysis is limited to radii $\leq 3$ Mpc, while the Caustic
$M(r)$ determination extends to $\sim$5 Mpc. Moreover, beyond $\rume$
the lensing $M(r)$ determination is affected by the presence of a
large-scale structure feature contaminating the cluster line-of-sight
\citepalias{Umetsu+12}.  An additional advantage of using the Caustic
$M(r)$ determination at large radii is that we do not rely on the NFW
parametrization, which might not provide an adequate fit to the mass
density profile of virialized halos much beyond their virial radius
\citep{NFW96}. Since the Caustic and lensing $\mtwo$ values are
consistent but not identical, we re-evaluate the Caustic $M(r)$ (and
its errors) starting from $\rume$ outwards, assuming the lensing
$\mtwo$ value at $\rume$.

The resulting mass profile is shown in Fig.~\ref{f:massprof} where we
also display the joint MAMPOSSt+Caustic NFW best-fit model for
comparison. It is the first time that it is possible to constrain the
$M(r)$ of an individual cluster from 0 to 5 Mpc (corresponding to $2.5
\, \rtwo$) with this level of accuracy.  In the next Section we will use
this mass profile to determine the orbits of different galaxy populations
within the cluster.

\section{The velocity anisotropy profile}
\label{s:beta}
In the previous Section we determined a fiducial mass profile (shown
in Fig.~\ref{f:massprof}) that we now use to determine the velocity
anisotropy profiles of different cluster galaxy populations, via
inversion of the Jeans equation, a problem first solved by
\citet{BM82}. In our analysis we solve the sets of equations of
\citet{SS90} and, as a check, also those of \citet{DM92}. Similarly to
what was done by \citet{BK04}, our procedure is almost fully
non-parametric, once the mass profile is specified. In particular, we
do not fit the number density profiles (at variance with what we did
in Section~\ref{ss:compl}), but we apply the LOWESS technique
\citep[see, e.g.,][]{Gebhardt+94} to smooth the background-subtracted
binned number density profiles. We then invert the smoothed profiles
numerically \citep[using Abel's equation, see][]{BT87} to obtain the
number density profiles in 3D. We use LOWESS also to smooth the binned
$\slos$ profiles.

Since the equations to be solved contain integrals up to infinity, we
need to extrapolate these smoothed profiles to infinite radius. In
practice we approximate infinity with $R_{\infty}=30$ Mpc and we
check that increasing this radius to larger values does not affect our
results.  We extrapolate the LOWESS smoothing of $n(R)$ beyond the last
observed radius, $R_l$, with the following function: 
\begin{equation}
 n(R)=\eta \, (R_{\infty}-R)^{\xi}/R^{\zeta}, 
\end{equation}
with
\begin{equation}
\begin{array}{l}
 \zeta=[{\rm d}\log n/{\rm d}\log R]_{R_l} - \xi \,
 R_l/(R_{\infty}-R_l), \nonumber \\
\eta=n(R_l) \, R_l^{\zeta}/(R_{\infty}-R_l)^{\xi}. \nonumber
\end{array}
\end{equation}
The only free parameter in the extrapolating function is the $\xi$
parameter. We extrapolate the LOWESS smoothing of $\slos$ beyond the
virial radius\footnote{Dynamical relaxation of the cluster may not
  hold beyond $\rtwo$, so we prefer not to use the kinematics of
  cluster galaxies at larger radii in the Jeans equation inversion.},
$\rume$, by assuming that $\slos$ at $R_l$ is a fixed fraction of the
peak $\slos$ value, and by making a log-linear interpolation between
$\log \rtwo$ and $\log R_l$.  The $\br$ solutions are rather
insensitive to different choices of the extrapolation parameters (any
change is well within the error bars -- see below).

The dominant source of error on $\br $ arises from the uncertainties
in $\slos$.  It is however virtually impossible to propagate the
errors on $\slos$ through the Jeans inversion equations to infer the
uncertainties on the $\br$.  We then estimate these uncertainties the
other way round. We modify the $\beta$ profile in a generic way as
follows, $\br \rightarrow \br + \varepsilon_1 + \delta_1 \, r$, and
$\br \rightarrow \varepsilon_2 \, \br + \delta_2$.  We then compute
the predicted $\slos$ profiles for all values of $\{\varepsilon_1,
\delta_1\}$ and $\{\varepsilon_2, \delta_2\}$ in a wide grid, using
the equations of \citet{vanderMarel94}. The range of acceptable $\br$
profiles is determined by a $\chi^2$ comparison of the resulting
$\slos$ profiles with the observed one.

\begin{figure}
\begin{center}
\begin{minipage}{0.5\textwidth}
\resizebox{\hsize}{!}{\includegraphics{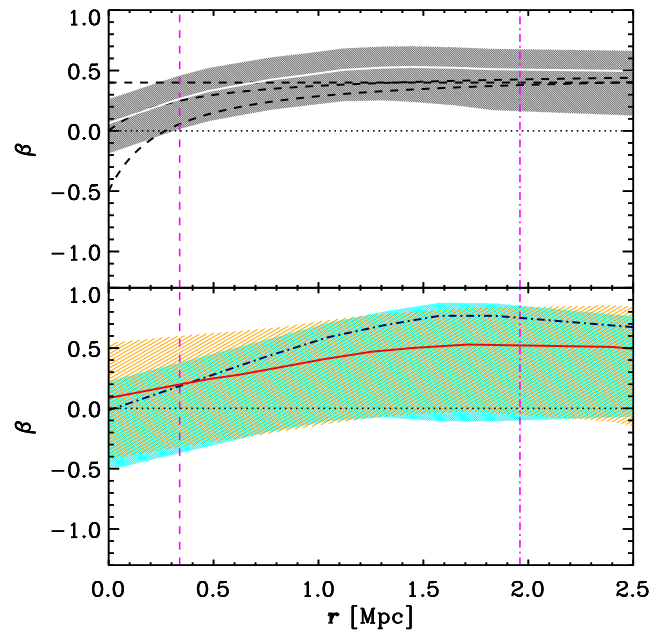}}
\end{minipage}
\end{center}
\caption{The velocity-anisotropy profile, $\br$, of different cluster
  galaxy populations.  {\it Top panel:} all cluster members.  The
  solid (white) curve is the solution of the inversion of the Jeans
  equation adopting the reference mass profile defined in
  Section~\ref{ss:combMr}. The hatched (gray) region indicates the
  1~$\sigma$ confidence region around this solution.  For comparison,
  three $\br$ models are shown (black curves).  They correspond to the
  best-fit $\br$ models of the MAMPOSSt analysis for a NFW $M(r)$
  model (see Sect.~\ref{ss:MAM}), namely (from top to bottom at small
  radii) the C, T, and O model.  In both panels, the vertical dashed
  and dash-dotted (magenta) lines indicate the location of $\rs$ and
  $\rtwo$, respectively, and the horizontal dotted line indicates
  $\beta=0$. {\it Bottom panel:} passive and SF cluster members,
  separately. The red solid (resp. blue dash-dotted)
  curve and orange (resp. cyan) hatched region
  represent the solution of the inversion of the Jeans equation within
  the 1~$\sigma$ confidence region for passive (resp. SF) cluster
  members. }
\label{f:beta}
\end{figure}

The $\br$ we obtain by this procedure using all cluster members is
shown in Fig.~\ref{f:beta} (top panel).  This is the highest-$z$
determination of an individual cluster $\br$ so far, and one of the
few available in the literature in a non-parametric form
\citep{BK04,Benatov+06,NK96,HL08,Lemze+11}.  It is isotropic near the
center, then it gently increases with radius, reaching a mild radial
anisotropy, $\beta \simeq 0.5$ at $\simeq \rtwo$. Constant, isotropic
velocity anisotropy is ruled out.

\begin{figure}
\begin{center}
\begin{minipage}{0.5\textwidth}
\resizebox{\hsize}{!}{\includegraphics{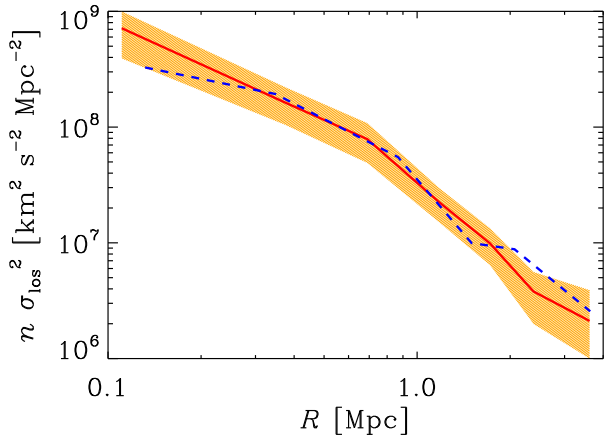}}
\end{minipage}
\end{center}
\caption{Consistency of the shapes of the $n(R) \, \slos^2(R)$
  profiles of the passive (solid red line) and SF (dashed blue line)
  cluster galaxy populations. The profile for the SF galaxy population
  has been multiplied by 3.7 to allow for a direct
  comparison with the profile of the passive galaxy population.  The
  hatched (orange) region indicates the 1 $\sigma$ confidence level of
  the profile of the passive population (that for the SF population is
  not shown, but it is much larger). }
\label{f:nslos}
\end{figure}

In Fig.~\ref{f:beta} we also display the best-fit $\br$ model obtained
by running the MAMPOSSt method with a NFW mass profile model (see
Sect.~\ref{ss:MAM}). All MAMPOSSt parametrized solutions are
consistent with this non-parametric determination over most of the
covered radial range. Note that the MAMPOSSt best-fit T $\br$ model is
identical to the model that has been shown \citep{ML05b,MBM10,MBB13}
to adequately describe the $\br$ of cluster-mass halos extracted from
cosmological simulations.

In the bottom panel of Fig.~\ref{f:beta} we show the $\beta$ profiles
of the passive and, separately, SF subsamples (defined in
Section~\ref{ss:members}). It is the first time that $\br$ is
determined separately for these two populations in an individual
cluster.  The two profiles appear very similar, and therefore also
very close to the $\br$ of all galaxies. Splitting the sample in two
clearly increases the error bars, so the passive and SF $\br$ are
formally consistent with isotropic orbits at all radii.

The remarkable similarity of the $\br$ of passive and SF galaxies may
seem unexpected given that their $n(R)$ are quite different (see
Fig.~\ref{f:nr}). However, the normalization of $n(R)$ is irrelevant
in the Jeans inversion equation and what matters is the combination
$n(R) \, \slos^2(R)$ (sometimes called 'projected pressure'), and
the normalization of $\slos(R)$. We have already seen that the values
of $\slos$ for the passive and SF cluster galaxy populations are quite
similar (see Table~\ref{t:slos}). In Fig.~\ref{f:nslos} we show that
also the shape of the $n(R) \, \slos^2(R)$ is rather similar for the
two populations, so the similarity of the passive and SF $\br$ is not
unexpected.

\section{$Q(r)$ and the $\gamma$-$\beta$ relation}
\label{s:qr}
With $M(r)$ and $\br$ we are now in the position to investigate the
$Q(r)$ behavior and the existence of the $\gamma$-$\beta$ relation
(see Section~\ref{s:intro}). It is the first time that these relations
are tested observationally in a galaxy cluster.  Both relations depend
on the mass density profile, $\rho(r)$, which is the same for all
tracers of the gravitational potential, but they also depend on other
quantities, the velocity dispersion and velocity anisotropy profiles,
which might in principle be different for different tracers. Clearly
we do not have access to these profiles for the DM particles, since
they are not observables\footnote{The derivation of $\br$ for DM
  particles done by \citet{HP07} is based on the strong assumption
  that the DM 'temperature' is identical to that of the hot
  intra-cluster gas at all radii, an assumption that cannot be
  verified observationally. A similar approach was followed by
  \citet{Lemze+11}, except that they used galaxies rather than
  intra-cluster gas for their derivation of the DM $\br$. Their
  approach is more appealing than that of \citet{HP07} because
  both DM particles and galaxies are collisionless, while gas is
  not.}, so we determine $Q(r)$ and the $\gamma$-$\beta$ relation
separately for different classes of tracers, namely all, passive, and
SF cluster members.

\begin{figure}
\begin{center}
\begin{minipage}{0.5\textwidth}
\resizebox{\hsize}{!}{\includegraphics{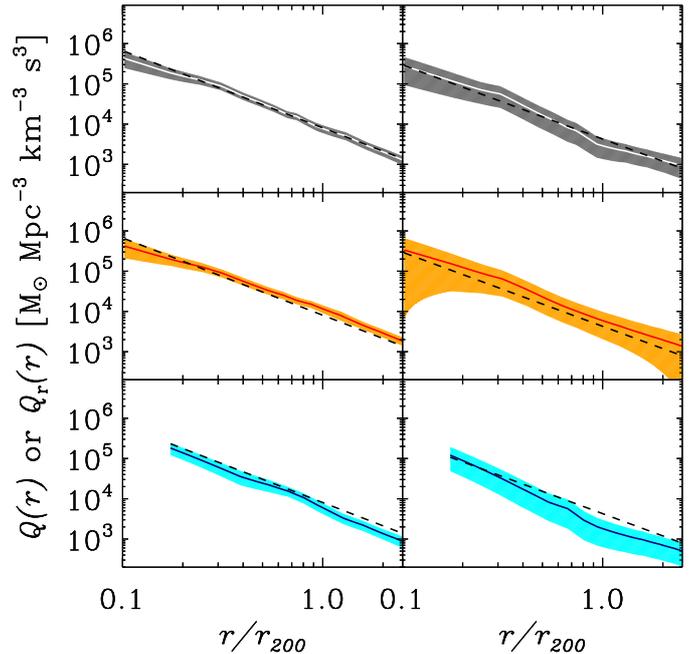}}
\end{minipage}
\end{center}
\caption{The pseudo phase-space density profiles $Q(r) \equiv
  \rho/\sigma^3$ (left panels) and $\qrr \equiv
  \rho/\sigma_{\rm{r}}^3$ (right panels), as a function of
  clustercentric radius $r$, within 1 $\sigma$ confidence regions
  (shaded area) for all (top panels), passive (middle panels), and SF
  (bottom panels) cluster members. The dashed lines are fixed-slope
  best-fit relations $Q(r) \propto r^{-1.84}$ and $\qrr \propto
  r^{-1.92}$ to the sample of all galaxies, where the slopes are those
  found by \citet{DML05} using numerically simulated halos.}
\label{f:qr}
\end{figure}

In Fig.~\ref{f:qr} we display $Q(r) \equiv \rho/\sigma^3$ (left
panels), 
and
$\qrr \equiv \rho/\sigma_{\rm{r}}^3$ (right
panels), for all, passive, and SF cluster members
separately.  The mass density profile $\rho(r)$ is obtained from our
fiducial mass profile (see Sect.~\ref{s:Mprof}) and $\sigma(r)$ and
$\sigma_{\rm{r}}(r)$ are obtained from the inversion of the Jeans
equation (see Sect.~\ref{s:beta}). The error bars are derived from the
uncertainties on $\rho(r)$ and $\br$, through a propagation of error
analysis; $\sigma_{\rm{r}}(r)$ is affected by much larger
uncertainties than $\sigma(r) $ because of the large
uncertainties on $\br$, i.e. we know the total velocity dispersion
better than we know its separate components.

In Figure ~\ref{f:qr} we also show the fixed-slope best-fit relations
$Q(r) \propto r^{-1.84}$ and $\qrr \propto r^{-1.92}$ using the sample
of all galaxies, where the slopes are those found by \citet{DML05} for
DM particles in numerically simulated halos. The sample of all members
obey both theoretical relations for $Q(r)$ and $\qrr$ within the error
bars.  Also the subsample of passive
members follows the theoretical relations, 
while the subsample
  of SF galaxies follows the theoretical relations only at $r/\rtwo
  \la 0.7$. 

\begin{figure}
\begin{center}
\begin{minipage}{0.5\textwidth}
\resizebox{\hsize}{!}{\includegraphics{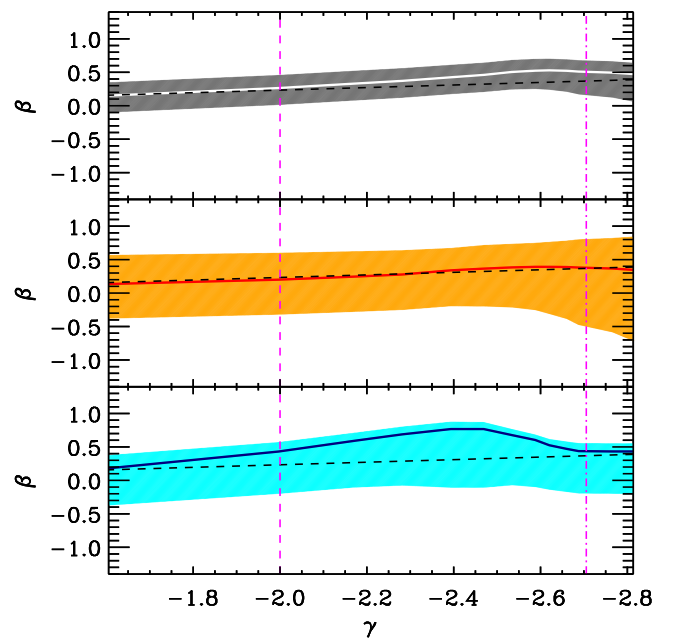}}
\end{minipage}
\end{center}
\caption{The relations between $\br$ and the logarithmic derivative of
  the total mass density profile, $\gamma(r)$, for all, passive,
  and SF member galaxies (top, middle, and bottom panel,
  respectively), within 1 $\sigma$ confidence regions (shaded
  regions), and the theoretical $\gamma$-$\beta$ relation of
  \citet{HM06} (dashed line). The vertical lines indicate the location
  of $\rs$ (dashed) and $\rtwo$ (dash-dotted).}
\label{f:gammabeta}
\end{figure}

In Fig.~\ref{f:gammabeta} we show $\br$ vs. the logarithmic derivative
of the mass density profile, $\gamma(r)$, for all, passive, and SF
members, separately, and the theoretical $\gamma$-$\beta$ relation of
\citet[][see eq.~\ref{e:betaHM}]{HM06}.  The theoretical relation is
consistent with the data within the observational error bars for the
full sample of members.  Passive galaxies obey the theoretical
$\gamma$-$\beta$ relation very well at all radii.  On the other hand,
the observed relation for SF galaxies deviates from the theoretical
one, especially at large radii, but this deviation is not
statistically significant, given the rather large observational
uncertainties.

\section{Discussion}
\label{s:disc}
\subsection{The mass profile}
\label{ss:mr}
Using a large spectroscopic sample of $\sim 600$ cluster members as
tracers of the gravitational potential we have determined the $M(r)$
of the $z=0.44$ MACS~J1206.2-0847 cluster to an accuracy close to that
reached by the combined strong+weak lensing analysis of
\citetalias{Umetsu+12}, and over a wider radial range, reaching out to
5 Mpc (corresponding to $2.5 \, \rtwo$). The determination of a
cluster $M(r)$ to such a high level of accuracy and over such a wide
radial range is unprecedented for this redshift.

For the $M(r)$ determination we have used two kinematics-based methods,
MAMPOSSt and Caustic.  This is the first application of the new
MAMPOSSt method to an observed cluster. MAMPOSSt allows to determine
$M(r)$ in the cluster virial region, where the Caustic method
suffers from systematics, and Caustic allows to determine $M(r)$
beyond the virial region, where MAMPOSSt is not fully applicable
because of possible deviations from dynamical equilibrium. The two methods
are therefore complementary.

The MAMPOSSt analysis indicates that the cluster $M(r)$ is best fitted
by the NFW or by the Einasto model, although we cannot reject any
of the other mass models we have considered, Hernquist, Burkert, and
SIS. The SIS model best-fit requires however a 
very small value of the core radius (see
Table~\ref{t:mamposst}). The Caustic analysis shows that the NFW model
provides a reasonable fit at least out to $\sim 2 \, \rtwo$.  Beyond
that radius the uncertainties in the Caustic $M(r)$ determination
become very large and constraints on the shape of the mass profile are
too loose (see Fig.~\ref{f:massprof}).

Previous analyses of cluster mass profiles traced by galaxy
kinematics have generally found good agreement with the NFW model
\citep[see the review of][and references therein]{Biviano08} as we
find for MACS~J1206.2-0847. The Burkert model was however found to
provide a somewhat better fit to the stacked $M(r)$ of the ENACS
data-set \citep{Katgert+98} by \citet{BS04} and cored models were not
excluded by the analysis of a cluster sample extracted from the 2dFGRS
\citep{Colless+01} by \citet{BG03}. \citet{BG03} have also found the
$M(r)$ slope to be consistent with that of NFW up to $\sim 2 \rtwo$;
beyond that radius, the slope may become intermediate between those of
the NFW and Hernquist models, according to the analysis of the CAIRNS
cluster sample \citep{Rines+03}. These previous results were based on the
combination or averaging of several cluster data-sets, since the individual
cluster statistics was insufficient to constrain $M(r)$, unlike in our case.

The best-fit NFW model obtained by combining the results of the two
kinematic methods (via a weighted average) is very close to the
best-fit NFW model obtained by the combined strong and weak lensing
analysis of \citetalias{Umetsu+12} (see Fig.~\ref{f:mproj}).  The
accuracy level we reach on the $M(r)$ parameters is close to that
reached by the combined strong and weak lensing analysis. There is
also a very good agreement with the $M(r)$ estimate within $\sim
\rfive$ obtained by \citetalias{Umetsu+12} using Chandra X-ray data.

The excellent agreement we have found between the
kinematically-derived $M(r)$, the $M(r)$ from lensing, and the $M(r)$
from X-ray indicates that our and \citetalias{Umetsu+12}'s
results are free from possible systematics. It also
indicates that MACS~J1206.2-0847 is dynamically relaxed.

The $\slos$-based $\rtwo$ estimate is also in agreement with our other
estimates (Table~\ref{t:NFW} and Fig.~\ref{f:cMr}). This constrains
the velocity dispersion of passive cluster members to be within $\pm 10$\%
of that of DM particles, in agreement with the results of numerical
simulations \citep[see, e.g., Fig. 8 in][]{Munari+13}.

Cluster concentrations may be affected by major mergers \citep{HRSH07}
and/or baryon cooling \citep{GKKN04,Duffy+10,RBEMM13} which tends
to steepen the $cMr$ \citep{Fedeli12}.  Early adiabatic compression of
galactic DM \citep{BL10} can increase the concentration.  Dynamical
friction acting on orbiting galaxies can pump energy into the diffuse
DM component and flatten the inner density slope \citep{ElZant+04},
and this flattening can be interpreted as a decrease in concentration
\citep{RPV07}.  The difference in the $cMr$ of relaxed and unrelaxed
halos in simulations suggests that the average effect of mergers on
concentrations is not very strong \citep[see, e.g.,][see also
  Fig.~\ref{f:cMr}]{DeBoni+13,BHHV13}. Baryonic processes appear to
have a stronger effect, as can be seen by comparing the $cMr$ of
\citet{DeBoni+13}, obtained on hydrodynamical simulations, and that of
\citet{BHHV13}, obtained on DM-only simulations (see
Fig.~\ref{f:cMr}).

The $M(r)$ of MACS~J1206.2-0847 has a concentration $\ctwo=6 \pm 1$,
slightly higher than the average for halos at the same $z$ and of the
same mass ($\mtwo=(1.4 \pm 0.2) \times 10^{15} \, \msun$) extracted
from cosmological numerical simulations \citep{DeBoni+13,BHHV13}, but
well within the scatter of the theoretical $cMr$ (see
Fig.~\ref{f:cMr}). The substantial agreement between the observed and
theoretically predicted concentrations argues against an alignment of
the cluster line-of-sight and major axis. This is also suggested by
the fact that the cluster appears somewhat elongated in projection
\citepalias{Umetsu+12}.

Our result for $\ctwo$ is consistent with others obtained from
analyses of the kinematics of stacked cluster samples, both at low-
\citep{KBM04,BS06,Lokas+06} and high-redshift \citep{BP09}.  The
analyses of the kinematics of individual clusters have found
concentrations both in line \citep{Lokas+06,RD06} and above the
theoretical expectations
\citep{LM03,Lokas+06b,WL07,Lemze+09,WL10,Abdullah+11}.

The concentration of cluster galaxies (both all and only the passive
ones) in MACS~J1206.2-0847 is smaller than that of the total
mass. Assuming that light traces mass would then lead to an erroneous
mass profile determination.  The concentration we find for the passive
galaxies, $\rtwo/\rn=3.1 \pm 0.7$, is close to the average found by
\citet{LMS04} for K-band-selected galaxies in nearby clusters,
$\ctwo=2.9$. The concentration of the {\em luminosity} density profile
of cluster galaxies is only $\sim 10$\% higher than the concentration
of their number density profile, indicating little evidence for mass
segregation.  The ratio of the concentrations of the total mass and
the passive galaxies is $1.8 \pm 0.4$, close to that found by
\citet{BP09} for a stack of nearby clusters (1.7), but much higher
than that found by the same authors for a stack of $z \sim 0.55$
clusters (0.4). Other studies have found this ratio to be closer to
unity \citep{Carlberg+97-mprof,vanderMarel+00,BG03,KBM04,Rines+04}.
Possibly the relative concentration of total mass and cluster galaxy
distribution is related to the assembly history of a cluster or to
dynamical processes affecting the survival of galaxies near the
center, such as merging with the central BCG or tidal
stripping. Extending the analysis presented in this paper to other
clusters may help understand the physical origin of the relative
concentrations of mass and galaxy distribution in clusters.

\subsection{The velocity anisotropy profiles}
\label{ss:br}
We have determined the velocity anisotropy profiles, $\br$, of passive
and SF members, separately, for the first time for an individual
cluster. This was done from the inversion of the Jeans equation, using
our best guess for $M(r)$, derived from the combination of the
best-fit NFW $M(r)$ from the lensing analysis of
\citetalias{Umetsu+12} within $\rume$ and the Caustic $M(r)$ outside.
MACS~J1206.2-0847 is the highest-$z$ cluster for which $\br$ has been
determined, and one of the few at all redshifts.

In our analysis we have assumed spherical symmetry. The analysis of
numerically simulated halos by \citet{Lemze+12} has shown that this
assumption has almost no effect on the determination of $\br$ within
the virial radius.

We have found that the $\br$ of all cluster members is consistent with
that of cosmological halos in numerical simulations \citep[][see
  Fig.~\ref{f:beta}, top panel]{ML05b,MBM10,MBB13}. It is not
consistent with isotropy at all radii, but only up to $\sim \rs$, then
it increases to more radial anisotropy.

The $\br$ for passive and SF cluster members are almost identical (and
therefore also almost identical to the $\br$ of all cluster members).
This is quite remarkable given that the two cluster populations have
different $n(R)$, i.e. they occupy different regions in the cluster.
However, the $\slos$ of the two populations are not significantly
different (see Table~\ref{t:slos}), and the $n(R)$ and $\slos(R)$ of
the two populations combine to produce $n(R) \, \slos^2(R)$ profiles
of similar shapes (see Fig.~\ref{f:nslos}). Hence the observable that
enters the Jeans equation inversion (by which we estimate $\br$) is
very similar for the two populations.

This common shape of the orbital distribution of cluster galaxies
could be the result of violent relaxation followed by smooth accretion
\citep{LC09}. Violent relaxation is expected to occur at higher
redshifts, and isotropize orbits, and therefore should concern the
more central cluster regions. Galaxies that were accreted by the
cluster after the end of violent relaxation, would retain their
slightly radial orbital distribution, producing the external
$\br$. Yet another process capable of isotropizing the initial radial
orbits of infalling galaxies is radial orbit instability \citep[ROI,
  see, e.g.,][]{Bellovary+08}.

To understand which is the physical process that shapes the orbits of
galaxies in clusters we must study the evolution of $\br$.  Most
previous observational determinations of $\br$ have been based on
stacked clusters or have been obtained by assuming a fixed model shape
of $\br$.  The whole cluster population has been found to move on
either isotropic \citep{vanderMarel+00,Rines+03,HL08}, or mildly
radial orbits \citep{Lokas+06} with a general increase of $\br$ from
nearly isotropic orbits near the center to moderate radial anisotropy
outside \citep{Benatov+06,Lemze+09,WL10}, similar to the profile we
find for MACS~J1206.2-0847.  The early-type, red, passive cluster
population has generally been found to move on isotropic orbits
\citep{Carlberg+97-equil,Biviano02,LM03,KBM04,HL08}, while the
late-type, blue, SF cluster population has been found to move on
slightly radial orbits \citep{BK04,HL08}. The $\br$ of SF galaxies in
the nearby clusters analyzed by \citet{BK04} is isotropic at radii
$r<\rtwo/2$, then it becomes more radial.

Comparison with the $\br$ of lower-$z$ clusters suggests that passive
galaxies undergo evolution of their orbits, more than SF galaxies, and
the orbits tend to become more isotropic with time.  Our result thus
confirms the suggestion of \citet{BP09}, which was based on a stacked
sample of clusters at $z \sim 0.5$ \citep[see also][]{Benatov+06}.
Since violent relaxation is a process that occurs on relatively short,
dynamical timescales, and at high $z$, one could argue that the
secular evolution of galaxy orbits toward isotropy is related instead
to a different process, possibly ROI. At variance with violent
relaxation, ROI continues even after the cluster has virialized
\citep{BWBD07}.  If the ROI timescale is long, this could explain why
we see orbital evolution for the passive cluster members, and not for
the SF ones, since SF galaxies would have the time to transform into
passive before ROI modifies their orbits.

Another process by which cluster galaxies could undergo orbital
evolution is via interaction with the ICM \citep{DBMS09}. Since this
process also quenches star-formation, it could naturally explain why
we observe $\br$ evolution for the quenched (passive) cluster
galaxies, and not for the SF ones. The timescale and importance of
this process needs however to be quantified to allow a more relevant
comparison with observational results.

Other results from numerical simulations are contradictory on the topic of
$\br$ evolution. \citet{Lemze+12} do not find significant evolution of
$\br$ with redshift.  \citet{Munari+13} finds that $\br$ for massive
clusters becomes mildly more radial at higher redshift. Their result
is consistent with that of \citet{Wetzel11} who finds that the orbits
of satellites at the moment of their infall within larger host halos
are more radial at higher $z$. On the other hand, \citet{ID12} find
the opposite redshift trend. 

To better understand the issue of $\br$ evolution, one needs a much
larger sample of clusters at different redshifts. There is
considerable variance in the shapes of the $\br$ of cluster-size halos
extracted from numerical simulations, even if located at the same $z$
\citep[see, e.g., Fig.1 in][]{MBB13}. Possibly, the $\br$ shape is
related to the shape of $M(r)$, and one cannot treat them
separately. Below, we discuss this point in detail.

\subsection{The pseudo-phase-space density profiles}
\label{ss:mrbr}
In Section~\ref{ss:br} we have argued for possible mechanisms capable
of explaining the $\br$ of different cluster populations. Combining
our knowledge of $M(r)$ and $\br$ can shed more light on this
topic. For the first time ever, we have determined $Q(r)$, $\qrr$, and
the $\gamma$-$\beta$ relation observationally, separately for all,
passive, and SF cluster members. All cluster members, and also,
separately, the subsamples of passive members, obey the
theoretical relations within the observational error bars (see
Fig.~\ref{f:qr} and Fig.~\ref{f:gammabeta}). Only for SF members
  there is some tension between the observed and theoretical
  relations, even if only at large radii, $\ga 0.7 \,
\rtwo$. 

\citet{DML05} have shown that, given the $\gamma$-$\beta$ relation and
the Jeans equation for dynamical equilibrium, $\qrr$ is a power-law in
$r$, with an exponent related to $\beta(0)$.  Based on their finding
\citet{DML05} argue as follows.  Violent relaxation would tend to
create a scale-invariant phase-space density (since the process is
driven by gravity alone), hence $\qrr \propto r^{\alpha}$. Dynamical
equilibrium would then force $\alpha$ to approach a critical value,
from which results the particular form of the $\rho(r)$ of
cosmological halos. A value $\beta(0) \simeq 0$ with radially
increasing $\br$ gives the $\alpha$ observed in numerical simulations.
The form of $\br$ could therefore result from the halo violent
relaxation followed by its dynamical equilibrium \citep{Hansen09}.

If this argumentation is correct, passive members of
MACS~J1206.2-0847 have undergone violent relaxation and have reached
dynamical equilibrium, while SF members seem to have not, although
  the current uncertainties are still rather large. Moreover, one
  would be tempted to conclude that baryonic processes are not
  particularly important in shaping the dynamical structure of galaxy
  clusters, since they are unable to change the $Q(r)$ of galaxies
  that have undergone violent relaxation.  

Comparison of the $Q(r)$ and $\qrr$ for a sample of clusters at
  different redshifts is needed for further insight. To our
knowledge, there is only another cluster for which a similar analysis
is being done (Munari, Biviano \& Mamon, in prep.). Also in this
nearby ($z=0.09$) cluster, the $Q(r)$ of red galaxies is in agreement
with the theoretical prediction, and that of blue galaxies is
not. The lack of evolution in the $Q(r)$ of passive galaxies is
  perhaps surprising, since SF galaxies become passive with time, and
  their $Q(r)$ is different from that of passive galaxies. Perhaps as
  SF galaxies get quenched, their $Q(r)$ evolves and approaches the
  theoretical prediction, but this would contradict the idea that
  $Q(r)$ is shaped by the process of violent relaxation alone.
  Another possibility is that the fraction of late-quenched galaxies
  in the spectroscopic data-sets of passive cluster members is small
  because late-quenched galaxies are fainter than the more pristine
  cluster passive members. This can happen because of downsizing
\citep[e.g.][]{NvdBD06}, or because of the effects of tidal
  stripping \citep[e.g.][]{Balogh+02}. Drawing conclusions on the
  basis of only two clusters is however premature. To shed more light
  on this topic $Q(r)$ must be determined for more clusters, over a
  range of redshifts, and for galaxies of different luminosities.

\section{Conclusions}
\label{s:conc}
We have analyzed the internal dynamics of the MACS~J1206.2-0847
cluster at $z=0.44$, based on a large spectroscopic sample of more
than 2500 galaxies in its field, mostly from VLT/VIMOS data
obtained in the context of the ESO large programme
186.A-0798. From this sample we have identified $\sim$600 cluster
members. This is the largest spectroscopic sample for cluster member
galaxies at $z>0.4$, and one of the largest available at any $z$.
Using this sample, we have applied the Caustic and, for the first time
on an observed cluster, the MAMPOSSt method, to determine the cluster
mass profile, $M(r)$.  These two methods do {\em not} rely on the
assumption that the spatial and/or velocity distributions of cluster
galaxies are identical to those of the DM particles.

We have found an excellent agreement between the $M(r)$ determined
using the projected phase-space distribution of cluster galaxies and
those determined by \citetalias{Umetsu+12}, who used a combined strong
and weak gravitational lensing analysis and Chandra X-ray data.  This
agreement indicates that possible systematic biases in our
dynamical analyses have been
  properly accounted for, and that the cluster is in a relaxed
dynamical state. The cluster $M(r)$ is best described by a NFW model,
but other mass profile models provide acceptable fits to our data.
The observed concentration of the best-fit NFW model is slightly above
current theoretical predictions, but not significantly so.  The
spatial distribution of all and passive cluster members is less
centrally concentrated than the total mass.  Using the velocity
dispersion of passive cluster members to estimate the cluster mass
gives a value in agreement with those obtained by the other, more
sophisticated, analyses. This suggests that the bias between the
velocity dispersion of passive cluster members and DM particles is
small, $\la 10$\%.

We have defined a fiducial $M(r)$ from the combination of those
obtained with the lensing and kinematic analyses, spanning a radial
range from the center to $\sim$5 Mpc (corresponding to $2.5 \,
\rtwo$). To our knowledge, this is currently the most accurate
determination of a cluster $M(r)$ over this radial range.  We have
used it to invert the Jeans equation and determine the velocity
anisotropy profiles, $\br$, for all cluster members, and, separately,
for passive and SF cluster members.  This is the highest-$z$
individual cluster for which $\br$ has been determined so far, and the
only one for which $\br$ has been determined separately for both
passive and SF galaxies.  We have found almost identical velocity
anisotropy profiles for the different cluster galaxy populations,
isotropic near the center (within $\sim \rs$) and increasingly
radially anisotropic outside. This profile resembles that of
DM particles in halos extracted from cosmological numerical
simulations. Comparison with nearby clusters suggests evolution
of the orbital profile of passive cluster members, but the physical mechanism
driving this evolution remains to be identified.

From the mass density profile and $\br$, thanks to the quality of our
$M(r)$ and the size of our spectroscopic data-set, we have been able
to determine the pseudo phase-space density profiles $Q(r)$ and $\qrr$
and the $\gamma$-$\beta$ relation. These are the first observational
determinations of these profiles and relation for a galaxy cluster.
They are consistent with the theoretical expectations 
in particular for the passive cluster
members.  This is probably an
  indication that these galaxies were in the cluster at the time of
  violent relaxation.  Marginal deviation from the theoretical
  relations is observed instead for the SF cluster members, suggesting
  that they are a more recently accreted population.

The cluster studied in this paper is part of a sample of 14 clusters
from the CLASH-VLT Large Programme with the VIMOS spectrograph, which
we expect to be completed in 2014.  In this paper we have shown that
with a spectroscopic sample of this size it is possible to constrain a
cluster $M(r)$ to an accuracy similar to that achievable by a
detailed, combined strong + weak lensing analysis.  It is also
possible to constrain the orbits of different cluster galaxy
populations in a non-parametric way by direct inversion of the Jeans
equation. Combining results from $M(r)$ and $\br$ it is possible to
test dynamical relations that inform us on the way cosmological halos
evolve and organize internally. We will extend this analysis to all
the CLASH clusters with sufficient spectroscopic coverage in the near
future, and this will allow us to explore the variance in the cluster
dynamical states, the $cMr$ for the total mass and the different
galaxy populations, and the universality of the $Q(r)$ and
$\beta$-$\gamma$ dynamical relations. Stacking dynamically-relaxed
clusters together could in the end even allow us to constrain the
equation of state of DM by comparison of the kinematically-derived and
lensing-derived mass profiles \citep{FV06,SDR11}.

\begin{acknowledgements}
We wish to thank Colin Norman for originally suggesting one of us (AB)
to determine the pseudo-phase-space density profiles. We also thank
Stefano Borgani, Gary Mamon, Giuseppe Murante, Tommaso Treu, and the
referee, Elmo Tempel, for useful suggestions and discussions, and Hans
B\"ohringer for providing phase 2 information on his VIMOS programme
(169.A-0595) also used in this study.  This research is partly
supported by the PRIN INAF 2010: "Architecture and Tomography of
Galaxy Clusters".  PR and IB acknowledge partial support by the DFG
cluster of excellence Origin and Structure of the Universe
(http://www.universe-cluster.de).  RD gratefully acknowledges the
support provided by the BASAL Center for Astrophysics and Associated
Technologies (CATA), and by FONDECYT grant N. 1130528.
\end{acknowledgements}

\bibliography{master}

\appendix
\section{The effects of different cluster membership definitions}
\label{s:syst}
The determinations of $M(r)$ and $\br$ described in Sect.~\ref{s:Mprof}
and \ref{s:beta} are based, at least in part, on the sample of cluster
members defined by the P+G procedure (see Sect.~\ref{ss:members}). Here
we examine how a different cluster membership definition affects our
results. For this, we here consider the membership definition obtained
with the Clean method instead of the P+G method. The two methods use
very different approaches for the identification of cluster members,
as described in Sect.~\ref{ss:members}.

In Table~\ref{t:diff} we list the fractional differences and
associated 1~$\sigma$ uncertainties of the $\rtwo, \rs$ and $\rn$
determinations obtained by using the two samples of cluster members
identified with the P+G and the Clean methods.  The effects of changing
the method of membership selection are marginal, as all changes are
within 1~$\sigma$. 

\begin{table}
\centering
\caption{The effects of changing the member selection method
(Clean vs. P+G).}
\label{t:diff}
\begin{tabular}{llc}
\hline 
Method & Sample & Clean/P+G  \\
\hline
& & \\  
\multicolumn{3}{c}{Quantity: $\rtwo$} \\
& & \\
From $\slos$ & $R \leq 1.98$ Mpc (passive only) & $1.07 \pm 0.07$ \\
MAMPOSSt  & $0.05 \leq R \leq 1.96$ Mpc  & $1.04 \pm 0.12$ \\
Caustic    & $R \leq 2 \times 1.96$ Mpc  & $1.03 \pm 0.22$ \\
& & \\  
\multicolumn{3}{c}{Quantity: $\rs$} \\
& & \\
MAMPOSSt  & $0.05 \leq R \leq 1.96$ Mpc  & $0.80 \pm 1.14$ \\
Caustic & $R \leq 2 \times 1.96$ Mpc   & $1.00 \pm 0.53$ \\
& & \\  
\multicolumn{3}{c}{Quantity: $\rn$} \\
& & \\
-- & $R \leq 1.96$ Mpc   & $0.96 \pm 0.20$ \\
--  & $R \leq 1.96$ Mpc (passive only) & $0.91 \pm 0.19$ \\
--   & $R \leq 1.96$ Mpc (SF only)  & $1.10 \pm 0.37$ \\
& & \\  
\hline
\end{tabular}
\tablefoot{We list the values of the ratios of several quantities,
$\rtwo, \rs, \rn$, obtained using the samples of cluster members
  identified by the Clean and P+G method, respectively. We also list
  1~$\sigma$ errors on these ratios.}
\end{table}

The $\rtwo$ estimates are all slightly increased when adopting the
Clean method instead of the P+G method, and this happens because of the
8 galaxies with high absolute values of $\vrf$ near the cluster
center selected as members by the Clean method but not by the P+G
method (see Fig.~\ref{f:rvm}). Since 7 of these 8 galaxies are passive,
the effects of the different membership selection are stronger on the
quantities derived using only passive galaxies.

The inclusion of these 8 galaxies in the sample of cluster members
causes a higher velocity dispersion estimate near the center, and
therefore a steeper $\slos$ profile. To accommodate for the steeper
$\slos$ profile near the center, the MAMPOSSt analysis forces more
concentrated mass profiles, with 20--25\% smaller $\rs$ estimates.
However, given the large uncertainties on the $\rs$ estimates these
changes are far from being significant. The Caustic $M(r)$ estimate is
less affected, because i) it is only partially based on the membership
selection within the virial radius, and ii) it uses all galaxies (and
not only members) also beyond the virial radius.

The $\rn$ estimates depend very little on which membership selection
is chosen, because i) they are based not only on the sample of
spectroscopic members but also on the sample of $z_{\rm{p}}$-selected
members, and ii) the inclusion of the 8 additional members near the
center has a smaller impact on $n(R)$ than it has on $\slos(R)$.

Given the marginal changes in the MAMPOSSt and Caustic estimates of
$\rtwo$ and $\rs$, using the Clean-based membership determination
instead of the P+G-based one, we still find consistency between the
$M(r)$ obtained via the MAMPOSSt and Caustic method and that of
\citetalias{Umetsu+12}. As a consequence, we would still adopt the
$M(r)$ of \citetalias{Umetsu+12} within $\rume$ and the Caustic $M(r)$ at
larger radii, and the resulting $M(r)$ would be almost identical to the
one we adopted using the P+G membership determination
(Sect.~\ref{ss:combMr}).

\begin{figure}
\begin{center}
\begin{minipage}{0.5\textwidth}
\resizebox{\hsize}{!}{\includegraphics{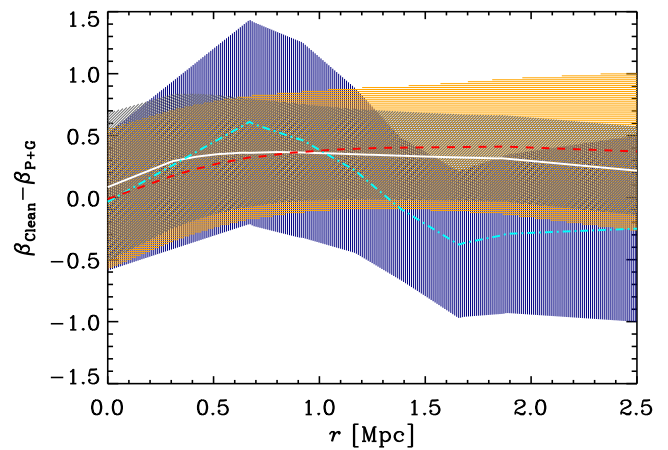}}
\end{minipage}
\end{center}
\caption{The difference of the $\br$ determined using the Clean
  and P+G samples of
  members.  The solid (white), dashed (red), and dash-dotted (cyan)
  curves are for all, passive, and SF galaxies,
  respectively. 1~$\sigma$ intervals on the differences are shown as shaded
  regions, with 45, 0, and 90 degrees orientation of the (gray,
  orange, blue) shading for all, passive, and SF galaxies,
  respectively. }
\label{f:betadiff}
\end{figure}

The $\br$ profiles resulting from the inversion of the Jeans equation
are marginally affected mostly because of the steepening of the
$\slos$ profile. Given that the adopted $M(r)$ is almost unchanged
with respect to the case of P+G membership selection, the steepening
of $\slos(R)$ near the center must be compensated by an increased
radial anisotropy.  This concerns mostly the passive galaxies. The
differences between the $\br$ obtained using the Clean-based sample of
members and those obtained using the P+G-based sample of members are
consistent with zero within 1~$\sigma$ for all cluster populations and
at all radii (see Fig.~\ref{f:betadiff}).

We conclude that our results do not change significantly if we use the
Clean instead of the P+G method for membership selection.

\section{Comparison with other cluster mass estimates from the literature}
\label{s:compMr}
We here compare our results to those
obtained by \citet{Foex+12} and \citet{Ebeling+09}. In both cases
their data were of insufficiently quality to constrain both $\rtwo$
and $\rs$, so we only compare the $\rtwo$ values.  

The weak lensing $\rtwo$ estimate of \citet{Foex+12},
$2.03_{-0.09}^{+0.11}$ Mpc, is in good agreement with our estimate.

\citet{Ebeling+09} have estimated the cluster mass in three ways; i)
by strong lensing, ii) by an hydrostatic equilibrium analysis of the
X-ray emitting intra-cluster medium, and iii) by the virial
theorem. Their strong lensing mass estimate, $1.12 \times 10^{14}
\msun$ within 0.12 Mpc from the cluster center, is in agreement with
our determinations. By applying a scaling relation to the cluster
X-ray temperature \citet{Ebeling+09} obtain an approximate
value of $\rtwo$, $2.3 \pm 0.1$ Mpc, in disagreement with our
estimate.  They then estimate the cluster mass within this radius
using an isothermal $\beta$ model profile, $1.7 \pm 1 \times 10^{15}
\msun$. This $\mtwo$ estimate corresponds to a $\rtwo$ estimate of 2.1
Mpc, different from their initial estimate, but still above our best
estimate. Had they iterated their eq.(5) they would have obtained a
concordant pair of $\rtwo, \mtwo$ estimates with a final value of
$\rtwo$ of 2.03 Mpc, closer to our best estimate.

The virial theorem mass estimate of \citet{Ebeling+09} is instead
grossly discrepant with any other estimate discussed so far. This
appears to be due to a combination of causes.  First, their membership
selection is too simplistic since it does not take into account the
radial position of galaxies. As a consequence, they obtain a much
larger velocity dispersion estimate than we do, 1581 \ks (compare to
the values in Table~\ref{t:slos}). Their large estimate is also due to
the fact that $\slos$ is decreasing with $R$ (see Fig.~\ref{f:slos})
and their spectroscopic sample does not reach $\rume$. Other causes
that lead \citet{Ebeling+09} to overestimate the cluster mass using
the virial theorem are the neglect of the surface-pressure term
\citep{TW86}, and the use of a spatially incomplete sample in the
estimate of the projected harmonic mean radius
\citep[see][]{Biviano+06}.

\end{document}